\documentclass[sigconf]{acmart}

\usepackage{orcidlink}

\AtBeginDocument{%
  }

\copyrightyear{2024}
\acmYear{2024}
\setcopyright{acmlicensed}\acmConference[CAIN 2024]{Conference on AI Engineering Software Engineering for AI}{April 14--15, 2024}{Lisbon, Portugal}
\acmBooktitle{Conference on AI Engineering Software Engineering for AI (CAIN 2024), April 14--15, 2024, Lisbon, Portugal}
\acmDOI{10.1145/3644815.3644951}
\acmISBN{979-8-4007-0591-5/24/04}

\begin{document}
\title{An Exploratory Study of V-Model in Building ML-Enabled Software: A Systems Engineering Perspective}

\author{Jie JW Wu  \orcidlink{0000-0002-7895-2023}}
\authornote{The author did this work after completing his PhD from George Washington University, before joining as a postdoc at the University of British Columbia, so there is no affiliated institution for this work.}
\affiliation{
\institution{}
\country{}
}
\email{jie.jw.wu@acm.org}

\renewcommand{\shortauthors}{Wu}

\begin{abstract}
Machine learning (ML) components are being added to more and more critical and impactful software systems, but the software development process of real-world production systems from prototyped ML models remains challenging with additional complexity and interdisciplinary collaboration challenges. This poses difficulties in using traditional software lifecycle models such as waterfall, spiral, or agile models when building \textit{\normalsize ML-enabled systems}. In this research, we apply a Systems Engineering lens to investigate the use of V-Model in addressing the interdisciplinary collaboration challenges when building ML-enabled systems. By interviewing practitioners from software companies, we established a set of 8 propositions for using V-Model to manage interdisciplinary collaborations when building products with ML components. Based on the propositions, we found that despite requiring additional efforts, the characteristics of V-Model align effectively with several collaboration challenges encountered by practitioners when building ML-enabled systems. We recommend future research to investigate new process models, frameworks and tools that leverage the characteristics of V-Model such as the system decomposition, clear system boundary, and consistency of Validation \& Verification (V\&V) for building ML-enabled systems. 
\end{abstract}

\maketitle

\section{Introduction}
Machine learning (ML) has been introduced to many software applications and has led to massive success~\cite{sarker2021machine}. However, it has been recognized that going from prototyped ML models to real-world production systems is challenging~\cite{akkiraju2020characterizing,amershi2019software,bosch2021engineering,shaw2022can}. Therefore, compared with traditional software systems, \textit{\normalsize ML-enabled systems} face additional challenges in the process~\cite{martinez2019crisp,haakman2021ai,serban2020adoption,studer2021towards}, data quality~\cite{sambasivan2021data, moller2020data, sambasivan2022deskilling, schelter2018automating},
technical debts~\cite{sculley2015hidden}, testing~\cite{zhang2020machine, riccio2020testing, braiek2020testing, huang2020survey}, security \& privacy~\cite{huang2011adversarial, liu2018survey, mcgraw2020architectural, wilhjelm2020threat}, safety~\cite{borg2019safely, salay2017analysis, salay2018using, zendel2015cvhazop},
and collaborations~\cite{nahar2022collaboration}, etc. Although the challenges and solutions of developing ML components have garnered considerable interest, such as engineering challenges~\cite{shaw2022can}, testing~\cite{braiek2020testing}, and automating development processes of ML components~\cite{braiek2020testing}, few studies have mentioned the interdisciplinary collaboration challenges~\cite{nahar2022collaboration,nahar2023meta} at system level for building product with ML components, including communication, documentation, engineering, and process~\cite{martinez2019crisp,haakman2021ai,serban2020adoption,studer2021towards}. In this work, we approach these challenges from a software process perspective.

Among the collaboration challenges, ML challenges the traditional software process lifecycle~\cite{braude2016software} due to its experimental and iterative nature. On one hand, with the introduction of ML into more and more impactful and critical software applications, collaboration problems start to emerge and be reported~\cite{nahar2022collaboration,nahar2023meta}, such as a lack of a standard process, unclear responsibility, and lack of a system-wide view, etc. On the other hand, the traditional software process lifecycle models such as waterfall, spiral, or agile models do not seem to fit well for ML-enabled system due to its different process compared with traditional software system~\cite{ozkaya2020really}. However, very few studies have explored new process models to systematically reduce the complexity, uncertainty, and risks of the software lifecycle for ML-enabled systems. The goal of this paper is to address this research question: \textit{\normalsize How can software process life cycles be established to address the interdisciplinary collaboration challenges from ML-enabled systems?}

\begin{figure*}[h]
\centering
\includegraphics[width=\textwidth]{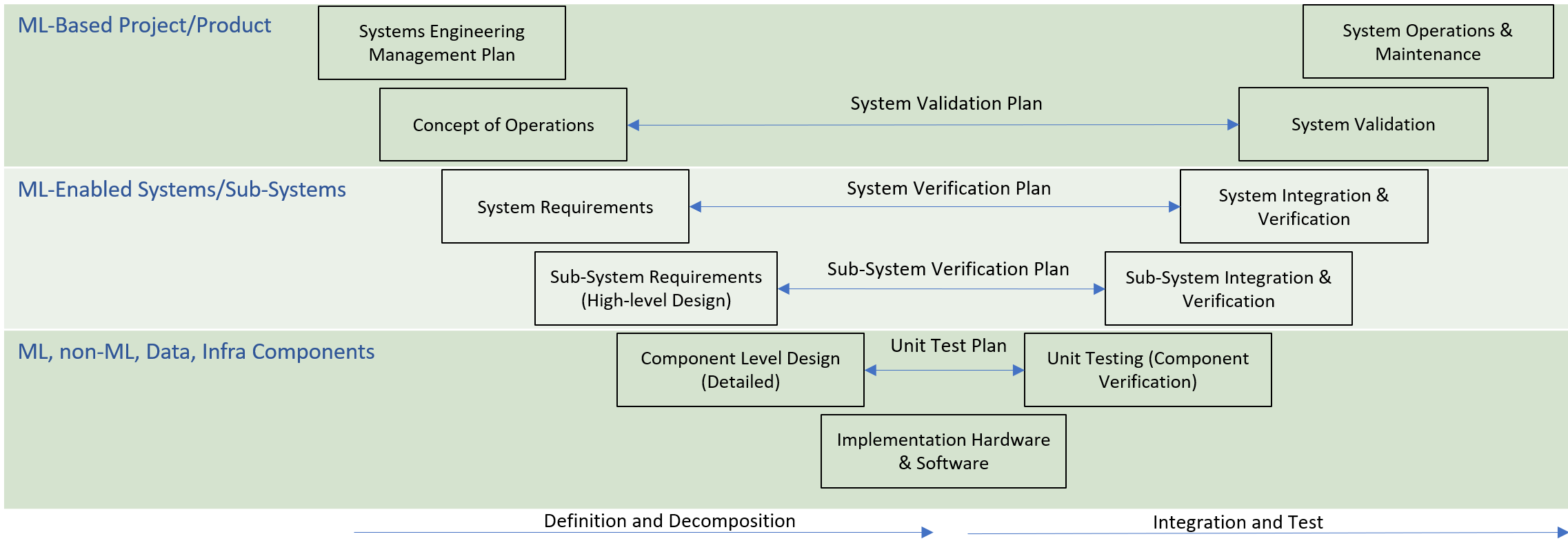}
\caption{V-Model~\cite{walden2015systems} in the context of building ML-enabled systems}
\label{fig:v4ml_vchart}
\end{figure*}

Systems Engineering (SE\footnote{SE stands for Systems Engineering rather than Software Engineering in this paper}) processes and principles provide a rigorous standard for large-scale projects with the goal of reducing complexity and risks~\cite{bodner2009handbook,sage2014handbook,Blanchard-Fabrycky-SystemsEngineering, Maier-Rechtin-ArtOfSystemsArchitecting}. Systems Engineering, an interdisciplinary field of engineering and engineering management, leverages system thinking principles to center on the problem of how to specify, design, integrate, and deliver successful complex systems over their life cycles~\cite{wiki:Systems_engineering}. The V-Model, originated in 1993 by ~\cite{brohl1993v} and commonly used in Systems Engineering processes~\cite{beale2006systems}, is an interdisciplinary, rigorous approach that starts from customer needs and requirements, then proceeds with design synthesis and system validation from system-level to sub-system level to component level, while taking care of problems such as performance, tests, and risks, etc. At its heart, ``V'' stands for decomposition on one thigh and integration on the other thigh, with continuous validation and verification (V\&V), as shown in Figure~\ref{fig:v4ml_vchart}. Motivated by the experience, interviews, and reflections of practitioners, we argue that the V-Model for Systems Engineering process, due to its interdisciplinary, system-wide, quality-focus nature, provides a way to address the literature gap in collaboration challenges in building ML-enabled software. Therefore, we consider the Systems Engineering process as a suitable lens to address our research question.

In this research, by interviewing practitioners from multiple software companies, we investigated the application of V-Model to address collaboration challenges in building ML-enabled systems. We analyzed the empirical evidence from the practitioners and derived a set of 8 propositions for using V-Model in building products with ML components. Our study had 11 participants who are professional software developers, ML scientists, and technical leaders. In the interviews, collaboration challenges were discussed for building ML-enabled systems, and then the use of V-Model in these situations was discussed. Our results from the propositions indicate that the V-Model takes additional efforts to follow, but they align well with the collaboration challenges encountered by practitioners when building ML-enabled applications.

\section{Background}

\subsection{ Software Engineering with ML Components}
We have seen an increasingly wide adoption of machine learning models in more and more critical and influential systems. Compared with traditional software products, building products with machine learning components is more science-like, uncertain, and probabilistic~\cite{smith2020machine}. This poses new challenges for traditional software lifecycle models~\cite{martinez2019crisp,haakman2021ai,serban2020adoption,studer2021towards}. Besides, a number of studies have discussed how to build better ML pipelines to train and deploy ML models, and new concepts are created, such as "MLOps"~\cite{makinen2021needs,serban2020adoption}. However, few studies focus on the system-wide process model that can better assist different roles such as developer, ML scientist, and leaders to work together with different components (ML, non-ML, data, infra). In this work, we explore the potential of finding a more suitable process model for building system-wide products with different roles and with both ML and non-ML components.

From the expertise perspective, ML-enabled systems impose expertise in a number of domains including data science, data quality~\cite{sambasivan2021data, moller2020data, sambasivan2022deskilling, schelter2018automating},
testing~\cite{zhang2020machine, riccio2020testing, braiek2020testing, huang2020survey}, security \& privacy~\cite{huang2011adversarial, liu2018survey, mcgraw2020architectural, wilhjelm2020threat}, safety~\cite{borg2019safely, salay2017analysis, salay2018using, zendel2015cvhazop}. These expertises come from different roles with different backgrounds and skill sets. Due to the introduction of ML components, these different roles, such as software engineer, ML scientist, technical leader, need to collaborate, and there are very few studies to address the challenges arising in these types of collaboration~\cite{nahar2022collaboration}. Regarding the lifecycle models for ML-enabled systems, ML lifecycle models such as CRISP-DM~\cite{martinez2019crisp} and Team Data Science Process (TDSP)~\cite{ericson2017team} were studied in the Fintech domain, with findings that ML lifecycle models need to be revised with more research~\cite{haakman2021ai}. In particular, these ML lifecycle models missed several necessary steps, including documentation, model evaluation, model monitoring, etc. Study that explored how teams develop, deploy and maintain software systems with ML components was conducted and summarized into best practices for improving software development processes ~\cite{serban2020adoption}. In this work, we attempt to address the collaboration problem from the perspective of a more interdisciplinary-focused process model.

\subsection{Systems Engineering Process and Principles}
Systems Engineering is a discipline closely related to Software Engineering. The former is focused on high-criticality and safety-critical systems. The latter is rapidly growing with more new methodologies and tools due to the flexible and malleable nature of software. Systems Engineering is a "full life cycle" engineering discipline that is responsible for starting and performing an interdisciplinary process to meet stakeholders' needs, reduce system complexity with high quality, well-defined schedule, and trustworthiness~\cite{haberfellner2019systems,bodner2009handbook,mathieson2020systems,beale2006systems}. The primary target of Systems Engineering process is the system that is complex and requires the expertise of different engineering disciplines. The goal is to reduce the system complexity, the risk of failures, cost overrun, and development time~\cite{Wasson-SystemsEngineeringCopingComplexity, Kossiakoff-SystemsEngineeringPrinciplesAndPractice,georgiadis2013using,bersson2012framework}. The hierarchy of the elements goes from system, to subsystem, to components, and to parts. Given the additional complexity and interdisciplinary collaboration that machine learning brings in, our motivation in this work is to investigate applying the V-Model from SE process to ML-enabled systems to address the challenges that have been raised~\cite{nahar2022collaboration,nahar2023meta,o2020common}.

\begin{table}[h]
  \caption{Comparison of lifecycle models on several dimensions~\cite{mathieson2020systems,cederbladh2023early,graessler2020new} in software and systems engineering. }
  \label{tab:model_comparison}
  \includegraphics[width=.48\textwidth]{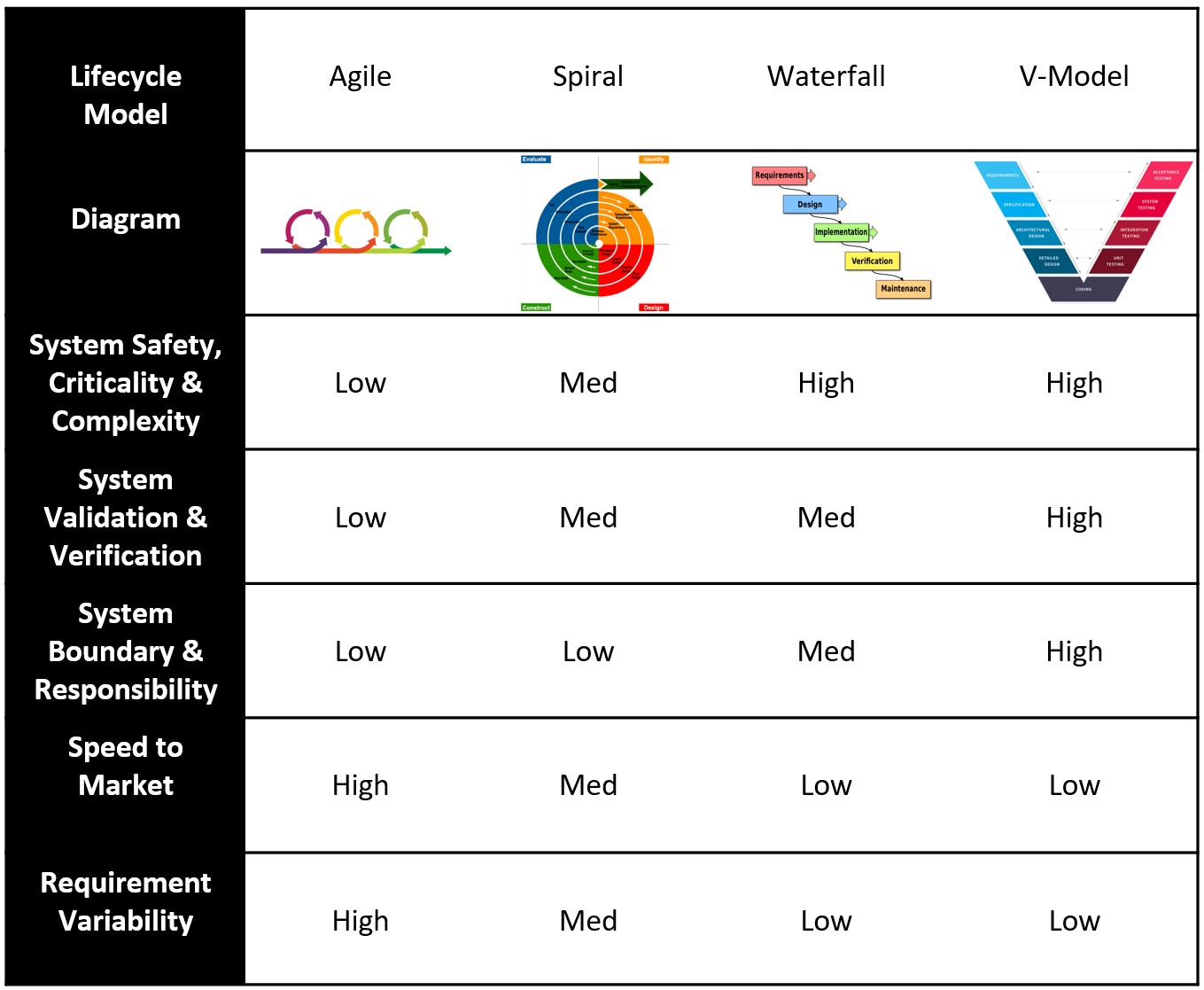}
\end{table}

From the software development lifecycle (SDLC) model perspective, different lifecycle models have different ways of balancing when and how to design interfaces. Table~\ref{tab:model_comparison} summarizes different mainstream lifecycle models with their advantages and disadvantages~\cite{mathieson2020systems,balaji2012waterfall}. In Table~\ref{tab:model_comparison}, the Waterfall model has a rigorously structured plan but with low variability. The Agile model, widely used in today's software development, focuses more on customer needs and "Speed to market" of features but is unstructured and poses challenges for complex systems such as ML-enabled systems~\cite{mathieson2020systems,balaji2012waterfall}. The Spiral model prioritizes iterative development and deployment for incremental delivery but has the risk of instability caused by changing requirements and also challenges with complex systems. Compared with Agile, Spiral, and Waterfall models, V-Model has disadvantages in "Speed to Market" and "Requirement Variability" given its limited flexibility~\cite{mathieson2020systems,balaji2012waterfall}, but has advantages in "System Boundary"~\cite{graessler2020new}, "System Validation \& Verification"~\cite{cederbladh2023early} for systems with high complexity such as ML-enabled systems. In this context, our research to investigate V-Model is about finding a proper balance for ML-enabled systems between the complexity, boundary, consistency, and criticality of systems, speed to market, and requirement variability. There have been recent research works that study the integration of SDLC models for machine learning applications~\cite{ranawana2021agile,laato2022integrating,petersen2022towards}, our work is different in that we mainly address the gap of collaboration challenges by developing a set of propositions through the use of V-Model.

\section{Research Methodology}
\subsection{Research Design Summary} The goal of the research is to explore the use of V-Model for building ML-enabled systems. Thus, we used an \textit{\normalsize explanatory approach} to develop a set of research propositions. The reasons we chose this approach are as follows: first, there are very few studies on the process lifecycle models for ML-enabled systems. Second, there is no prior knowledge of the application of V-Model to ML-enabled systems. Third, this approach is qualitative and thus provides a testable way to develop propositions by using induction from observations~\cite{bitektine2008prospective}. We briefly describe the interview and analysis procedure as follows. 

\subsection{Participants} Interviews were conducted with 11 participants, coming from 9 companies. The participants were hired from the author's two-hop connections in the software industry. Four of them work at Big Tech companies. The rest seven of them worked at mid-sized tech companies. Four of them had experience as ML scientists, and the rest mainly worked as software developers. Four of them had additional experience in technical leadership and management. All of them had more than 4 years of professional experience as either ML scientists or software developers. Their team had worked on at least one machine learning-based project. 

\subsection{Interview Protocol} We developed a semi-structured interview protocol as the interview guide, which includes a set of challenges and collaboration problems in the existing literature for ML-enabled systems~\cite{nahar2022collaboration,nahar2023meta}. The interview guides were sent to the interviewees ahead of time to give enough time for them to prepare the answers. Each interview took between 60 and 120 minutes, as some participants provided more details in some questions due to their practical experience and reflective insights in these questions.

\subsection{ Data Analysis} We used framework analysis~\cite{ritchie2002qualitative} due to the involvement of professional practitioners. The framework analysis includes five key stages: 1) familiarization, 2) identifying a thematic framework, 3) indexing (participants were assigned to \textbf{P1-P11}), 4) charting, 5) mapping and interpretation~\cite{ritchie2002qualitative}. We started the analysis after each interview was completed. This ensured we made small changes to the next interviews according to the valuable feedback from each interview. After the interviews were completed, for some questions that hadn't reached a consensus, we followed up with the interviewees to further discuss and validate our updated findings. Interviewees replied with comments and some additional findings.

\begin{figure*}[h]
  \centering
  \includegraphics[width=\textwidth]{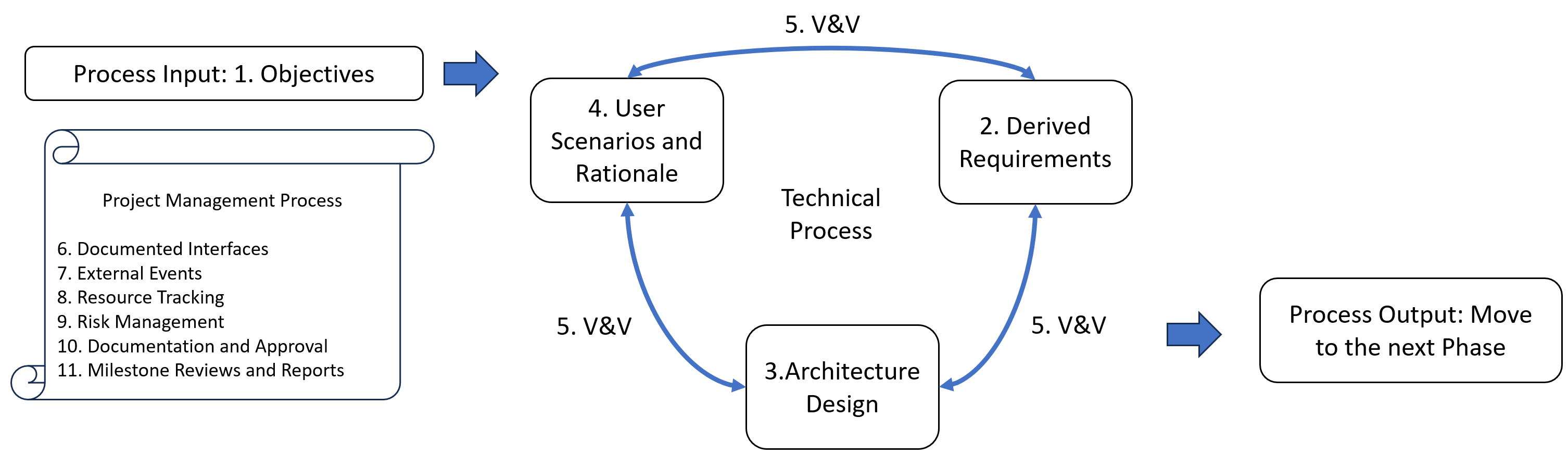}
  \caption{11 SE functions~\cite{beale2006systems} in the context of building ML-enabled systems}
  \label{fig:v4ml_functions}
\end{figure*}

\subsection{ Threats of Validity} Typical threats from qualitative research work also exist in this work. For example, the collected data may be different in companies with specialized applications, or geographic locations. The questions asked in the interview cannot possibly cover all challenges in ML-enabled systems. We mostly only interviewed one person for each company, which may be biased. No roles were interviewed outside the role of developer and scientist, except that only a small number of participants have leadership/management roles. The biases of authors in selecting the participants and other decisions in the interview protocol may affect the results, although we tried our best to follow the standards.

\section{V-Model Concept for Building ML-enabled Systems}
In this section, we describe the concept of Systems Engineering V-Model and put V-Model in the context of building ML-enabled systems. In a nutshell, the formula of Systems Engineering is 
\begin{equation}
\underline{\textit{\normalsize SE = Vee + 11 SE Functions + Tools}},
\end{equation}
in which the 11 SE functions are summarized in~\cite{beale2006systems}. Following the notation in~\cite{beale2006systems}, the 11 SE functions are donated as \textbf{SE1-SE11} in later sections, and Vee is donated as \textbf{V} in the propositions of the result section. We changed some of the original terminologies in the V-Model to better adapt to the context of ML Engineering. We describe Vee and 11 SE functions in detail below.

\subsection{Vee Chart}
\textbf{Definition.} Figure~\ref{fig:v4ml_vchart} shows the Vee chart and its context for building ML-enabled systems. As shown in Figure~\ref{fig:v4ml_vchart}, the Vee chart is the core construct of the V-Model. The Vee chart is composed of the following phases~\cite{cechini2009systems}:

\begin{itemize}
\item Pre-Phase A: Concept Exploration and Benefit Analysis
\item Phase A: Concept Development (System level)
\item Phase B: System Requirements (System level)
\item Phase C: High-Level Design and Sub-system Requirements (Subsystem level)
\item Phase D: Detailed Design (Component level)
\item Phase E: Verification \& Validation, Operations and Maintenance, Changes \& Upgrades 
\item Phase F: System Retirement / Replacement / Cleanup 
\end{itemize}

At a high level, in each phase of the Vee chart on the left leg, we apply 11 SE functions, to achieve the objectives. From the process perspective, given the target system or project of interest, this process first follows the left leg downwards, and then follows the right leg upwards, to finish each phase in order. The left leg is focused on formulation, with the goal of decomposing the system into subsystems and components and finishing the design. The right leg is focused on implementation, with the goal of integrating the system from the component level with verifications. Note that Pre-Phase A and Phase F are not explicitly shown in Figure~\ref{fig:v4ml_vchart}.

\textbf{Connection to ML Engineering.} One appealing feature of the Vee chart is the clear system boundary and responsibility between the system, subsystem, and components. We will show later that this addresses several challenges as mentioned by practitioners' evidence and propositions. Specifically, following the left leg, in Pre-Phase A, concept exploration and opportunity/benefit analysis are performed during the project planning to generate candidate ideas and select the project based on the opportunity sizing data or stakeholders' needs. In Phase A, how the system operates to achieve stakeholders' needs is described. In Phase B, requirements at the system level are created, followed by the high-level design for the system and the sub-system requirements in Phase C. Note that in ML development, Phase B and Phase C could be merged into a single Phase if there is no need to distinguish system and sub-system. Phase D is the detailed design at the component level for ML, non-ML, data, infra components, etc.

On the right leg, Phase E is building up components and subsystems to form the system, with a focus on testing at both the component level and system level, where the issue of unclear responsibility for system-level testing is stated by both literature~\cite{nahar2022collaboration} and practitioners. Finally, Phase F explicitly calls out the system retirement, replacement, or cleanup, which is reported by practitioners to be missing for several ML-based projects.

\begin{table*}
  \caption{11 Systems Engineering functions, with original function names, adapted names in the context of ML Engineering, and their brief descriptions~\cite{beale2006systems}}
  \label{tab:sefunc}
  \begin{tabular}{cp{3cm}p{3cm}p{8cm}}
    \toprule
    \textbf{Identifier} & \textbf{Original SE Function Name} & \textbf{Adapted Name for ML Engineering} & \textbf{Description} \\
    \midrule
    SE1 & Mission Objectives and Constraints & Objectives & Statements describing the goals and constraints of the (mission) objectives \\
    \midrule
    SE2 & Derived Requirements Development &Derived Requirements Development & Concise statements outlining 1) what to achieve, 2) how the desired level of achievement, and 3) any associated constraints or limitations.\\
    \midrule
    SE3 & Architectural Design Development & Architectural Design Development& Explanation of elements and interfaces, with detailed analysis for assessing the performance \\
    \midrule
    SE4 & Concept of Operation (“ConOps”) &  User Scenarios and Rationale& Explanation of how the system operates under the (mission) objectives to align with stakeholder expectations \\
    \midrule
    SE5 & Validate and Verify & Validate and Verify & An ongoing SE function, ensuring consistency in requirements, ConOps, and architectural design during formulation, while also planning and executing tests under a verification plan \\
    \midrule
    SE6 & Interfaces and ICD (Interface Control Document) & Documented Interfaces& Descriptions and documentations of interfaces, boundaries between elements, that evolve as the architectural design with requirements outlined. \\
    \midrule
    SE7 & Mission Environment & External Events & Descriptions and documentations of external environmental concerns and exposures \\
    \midrule
    SE8 & Technical Resource Budget Tracking & Resource Tracking& Tracking and identification of resource budgets \\
    \midrule
    SE9 & Risk Management & Risk Management& Negative occurrences and their potential unfavorable impacts on the project  \\
    \midrule
    SE10 & Configuration Management and Documentation &Documentation and Approval & A system for documentation control, access, approval and dissemination \\
    \midrule
    SE11 & System Milestone Reviews and Reports & Milestone Reviews and Reports& Reviews that communicate to stakeholders with significant project events, decision points, and achievements, etc.
 \\
  \bottomrule
\end{tabular}
\end{table*}

\subsection{11 SE Functions}
For each Phase of the Vee chart, 11 SE functions are applied to achieve its goals. Table~\ref{tab:sefunc} displays the original names and detailed definitions of the 11 SE functions. For each of the 11 SE functions, we describe and discuss the connection to ML Engineering below. Note that for each SE, if there is a modification in terminology due to adaptation to ML engineering, we include the original terminology within brackets. The 11 SE functions and their context for building ML-enabled systems are visually illustrated in Figure~\ref{fig:v4ml_functions}.

\subsubsection{SE1: Objectives (Mission Objectives and Constraints)}\hfill\\

SE1 refers to the objectives or goals in the context of ML Engineering. A similar mapping of SE1 in the software management process is Objectives and Key Results (OKR)~\cite{niven2016objectives}, a goal-setting method commonly used in software companies to set measurable goals.

\subsubsection{SE2: Derived Requirements Development}\hfill\\

SE2 is about requirement analysis that converts the user scenarios and stakeholder objectives into requirements. Categorization of requirements includes functional, performance, verification, and interface requirements. Each level has a separate set of requirements for that level (system, subsystem, component level). 
Based on practitioners' evidence, having requirements at the system level and component level would help address the challenge points in requirement engineering and model development, as illustrated in the reasoning of propositions PR1, PR4.

\subsubsection{SE3: Architectural Design Development}\hfill\\

SE3 captures the architectural design that describes the elements, interfaces, and expected performance based on requirements. According to the practitioners, given the complexity of the ML component, it is especially important to pay closer attention to the interfaces of the ML components to other components and other systems/subsystems. Meanwhile, it's also critical to have design at the system level to avoid potential system-wide issues caused by ML components, as mentioned in proposition PR2.

\subsubsection{SE4: User Scenarios and Rationale (Concept of Operation)}\hfill\\

SE4 mainly corresponds to the role of the product manager and designer. It is about how the system will be used from the end-user and customer perspectives. It represents the customer needs that lead to system requirements. One practitioner had experienced that when working on an ML-based project, the software engineer was focused totally on the interaction and development of ML, infra, and data components on the server side, but missed a small improvement on the UI side, which later turned out to be very impactful from the user engagement perspective (P1). So this part of the process is essential.

\subsubsection{SE5: Validate and Verify}\hfill\\

SE5 captures the Verification \& Validation (V\&V) between requirements, architectural design, and user scenarios, at both the design synthesis stage (left leg of Vee) and system validation stage (right leg of Vee). Verification is \textit{"assuring that the system is built right"}~\cite{fanson2010lessons}. Validation is \textit{"assuring that the right system is built"}~\cite{fanson2010lessons}. On the left leg side, V\&V is a function to ensure that the design, requirements, and user scenarios are consistent with each other. On the right leg side, V\&V makes sure that testing is executed at the system, subsystem, and component level, which is reported as an unclear responsibility. 
Most practitioners mentioned that in data-driven organizations, system validation refers to the validation of the impact of ML-based projects on moving the key metrics. This is typically done via online evaluations such as A/B testing to make data-driven decisions~\cite{wu2023comparison}. Practitioners agreed that testing and online evaluation as V\&V are critical (proposition PR6). Besides testing and online evaluation, some practitioners reported inconsistency between artifacts or understanding of design synthesis and actual implementation (P1,P2), which can be also addressed via V\&V.

\subsubsection{SE6: Documented Interfaces (Interfaces and Interface Control Document)}\hfill\\

SE6 refers to the documented interfaces in the design. They can be either internal or external interfaces. Since the interface is a key factor for better collaboration, the definition of interfaces should be a focus of the entire process. 
One practitioner mentioned that having a clear contract (or interface) with other components and other roles is very important when working on ML-based complex projects (P8). Interfaces of components such as data and infra should be especially emphasized, as mentioned in proposition PR5. Once there is any change on either internal or external interfaces that leads to project delays, we should actively update the related artifacts and schedule in the project process.

\subsubsection{SE7: External Events (Mission Environment)}\hfill\\

SE7 captures any external event at the organization level or team level that affects the project's progress. 
One practitioner had experienced that the feature launches and feature work were paused due to a team-wide investigation on metrics (P1). Another example is that the design is changed to add a constraint on the cost budget due to an external event of company-wide cost tracking. Several practitioners also mentioned that urgent investigations and on-call issues required non-trivial efforts and analysis since the ML complexity had an impact on their ongoing projects.

\subsubsection{SE8: Resource Tracking (Technical Resource Budget Tracking)}\hfill\\

SE8 refers to any resource tracking for the current project or ML-enabled systems. 
In the interviews, for ML-enabled projects, the cost of the existing jobs and services is one of the commonly mentioned resources. Another resource is the number of machines one can use for running model training jobs or data-related jobs. Regarding any pain point, one practitioner reported the uncertainty of infra jobs and ML model training jobs that caused periodic cost investigation in their team (P1).

\subsubsection{SE9: Risk Management}\hfill\\

SE9 captures any changes and the corresponding risks in the process such as requirement change or design change. 
Due to the uncertainty and complexity of the ML model, this is not uncommon and in part caused several challenge points according to our interviews with the practitioners (proposition PR3). With the risk management in the project process, any risk in the ML-enabled system would be first identified, and then evaluated. Finally, the corresponding method will be developed as mitigation to the risk.

\subsubsection{SE10: Documentation and Approval (Configuration Management and Documentation)}\hfill\\

SE10 refers to the documentation control, access, and approval. 
According to the practitioners, having documentation at different levels, and keeping them up-to-date is helpful, although it needs a certain amount of work (proposition PR8). As also mentioned in the existing literature~\cite{haakman2021ai,nahar2022collaboration,nahar2023meta}, more research is needed in documentation to ensure a good trade-off is obtained between efforts on documentation and its benefits. 

\subsubsection{SE11: Milestone Reviews and Reports}\hfill\\

SE11 captures the reviews and artifacts in several critical milestones of the ML-based project. Most practitioners acknowledged the importance of design review, as it provides a way to communicate with both the external and internal stakeholders and collect important feedback on the design.

\section{Results on Challenge Points}
The challenge points of ML-enabled systems are discussed in the interviews with the practitioners from software companies. Most of these challenges are concerned with interdisciplinary collaboration challenges. For each area, we briefly mentioned the challenges from the existing literature and then discussed the related empirical evidence from the practitioners. For the literature, we mainly followed~\cite{nahar2023meta,nahar2022collaboration} that provided a comprehensive summary and analysis of the challenges. Finally, we derived a set of propositions (\textbf{PR} for short) regarding how the process should be enacted from the perspective of the Systems Engineering process.

\subsection{Requirement Engineering}
\subsubsection{Challenges in Literature}
The existing literature mentioned challenges in requirements such as vague specifications in ML problems and unrealistic expectations from managers and other team members due to a lack of AI competence~\cite{nahar2023meta}. The vagueness in ML problem specifications refers to the difficulty in mapping between the high-level business goals and low-level requirements of the ML model, and measuring the contributions of the ML model ~\cite{nahar2023meta,haakman2021ai}. The unrealistic expectations exist in the customers, and team members: the customers' view of model development is static, and team members had difficulties in defining and scoping the projects~\cite{nahar2023meta,namvar2023beyond}. 
\subsubsection{Discussions from Practitioners}
For vagueness of ML specifications, a number of practitioners acknowledged that due to the experimental nature of ML models, it's nearly impossible to map the changes in ML components directly and accurately to changes in terms of business metrics or performance metrics (P1,P6,P8,P9,P10). Some practitioners from big tech companies mentioned that in their design reviews before implementation, they typically would list the requirements to provide an estimate of the change in business metrics or performance metrics (P1,P8,P9). In this way, support from the infra side can be planned and discussed if the change will cause relatively big regressions on performance metrics such as latency or Queries per second (QPS). Some practitioners reported that they lack such type of requirements in their projects (P2,P3,P4). For instance, one practitioner shared that \textit{\normalsize "at first, we just wanted to run the ML model when users enter the full sentence and click SAVE, but we found the ML model had very good results, so we changed the requirement to run the model each time when the user adds a new word. This significantly increased the QPS at launch time, and several engineers and on-calls were pulled in to investigate this urgent issue until it was resolved."} (P2) This caused the operational on-call issue that the QPS of the ML model was unexpectedly high during the first day of launch. The practitioner agreed that this problem could be avoided if the functional requirements and technical solutions of different alternatives had been created and reviewed in the design doc.

As for the unrealistic expectations, one practitioner mentioned that their leaders did not have a detailed level of understanding of ML and AI (P5). This has caused unrealistic expectations from them on the quarterly metric goal that can be driven by ML models. To set the right expectation, the engineers educated the leaders by giving a number of concrete examples to demonstrate the upper limit of how ML can help with the ranking problem in their team. Overall, most practitioners agreed that listing the requirements at the system level for ML-based projects helps to avoid potential issues and unrealistic expectations of requirements. This evidence led to the following proposition:
\hfill\\ \\
\fbox{\begin{minipage}{25em}
\textbf{\normalsize PR1}: System-level requirements should be created and actively maintained to keep up-to-date with new requirement changes, with the participation of the owners of ML components and non-ML components. (V, SE2, SE5, SE9)
\end{minipage}}

\subsubsection{Contributions of the V-Model}
System system-wide view is often lacking when building products with ML components~\cite{nahar2022collaboration}. The V-Model brings in clear system boundaries and responsibility between systems, subsystems, and components. This leads to the necessity of requirements at the system(or subsystem)-level in PR1. Besides, the risk management (SE9) and V\&V (SE5) of V-Model ensure the requirements in PR1 are actively maintained and up-to-date. This can be easily missed and cause issues when uncertainty and complexity are high in building ML-enabled systems.  

\subsection{Architecture, Design, and Implementation}

\subsubsection{Challenges in Literature}
In terms of architecture, design, and implementation, the literature highlighted the difficulty of the transition from model-centric to system-wide view in production~\cite{nahar2023meta,nahar2022collaboration}. Meanwhile, several studies reported the delay in planning and monitoring of change~\cite{nahar2023meta,nahar2022collaboration,lewis2021software,amershi2019software,arpteg2018software}. From software design and architecture perspective, studies have found additional complexity in system design for incorporating ML models, typically caused by dependencies, and entanglements~\cite{nahar2023meta,dove2017ux,lewis2021software,li2022testing}. 
\subsubsection{Discussions from Practitioners}
Let us categorize the practitioners into two groups: in the first group, the practitioner does both the ML role at the ML model level and the software developer role at the product level as a single engineer. Therefore, for this group, there is no issue of such a transition, although one practitioner argued it's a \textit{\normalsize "heavy responsibility"} that many different things with different contexts are done by one person.

In the other group, software developers collaborate closely with ML engineers to deliver ML-based projects. For this group, challenges of the transition do exist as reported in the interviews. Several practitioners had the experience that after the ML model was developed and evaluated by an ML engineer, the ML model was put in production, and issues such as regressions in performance metrics such as latency, and infra problems caused by QPS increase were later identified (P1,P2,P8,P9). Additional resources from software engineers or infra engineers were pulled into investigating these issues.

The practitioners mostly agreed that having a well-thought-out architecture design at the system level with clear interfaces, responsibilities, alternatives, and estimated performance helps alleviate the transition issues. Some practitioners argued that although providing an estimate generally helps, it's very difficult or unrealistic to have an accurate estimate during the design. There are many assumptions in the design phase, so how the ML model impacts the business and performance metrics can only be known when it's implemented and tested in production. In such cases of unexpected changes, as part of risk management (SE9), it's also helpful to update the design, any artifact, and planning as mitigation to avoid outdated information, which sometimes happens and misleads other team members. This evidence led to the following two propositions:
\hfill\\ \\
\fbox{\begin{minipage}{25em}
\textbf{\normalsize PR2}: System-level architecture design with elements, interfaces, responsibilities, alternatives, and expected performances should be created and actively maintained to keep up-to-date with new changes, with the participation of the owners of ML components and non-ML components. (V, SE3, SE5)
\end{minipage}}
\hfill\\ \\ \\
\fbox{\begin{minipage}{25em}
\textbf{\normalsize PR3}: Risks such as design changes or improvements due to uncertainty in ML components must be actively identified and mitigated. (SE5, SE9)
\end{minipage}}

\subsubsection{Contributions of the V-Model}
Similar to requirement engineering, the V-Model brings in 1) clear system boundary and responsibility, and 2) enforcement of V\&V, risk management that facilitates PR2 and PR3 in the context of building ML-enabled systems with high uncertainty and complexity. However, for PR3, one limitation of V-Model is its limited agility and adaptability to changes~\cite{balaji2012waterfall}, as pointed out in Table~\ref{tab:model_comparison}. As ML development is iterative and experimental in nature, if there is any change during the project, all the related documentation will have to be updated to the new change, leading to a slow top-down process. More future research is needed to address the rigid and inflexible nature of the V-Model to be adopted in organizations for iterative ML development. 

 \subsection{Model Development}

\subsubsection{Challenges in Literature}
In model development, the existing literature reported the conflicts in code quality due to a lack of standardized code quality in model development tools~\cite{nahar2023meta,siddik2023code,wan2019does}, and limited infrastructure and technical support~\cite{nahar2023meta,arpteg2018software}.

\subsubsection{Discussions from Practitioners}
For model development, several practitioners agreed on the low code quality of ML engineers given their educational backgrounds are more on science and statistics~\cite{kim2017data,nahar2022collaboration}. For example, P6 shared that \textit{\normalsize "SDEs (Software Development Engineers) are more emphasized on the coding standards. Scientists are more focused on model accuracy and are less focused on coding standards and code comments, class interface definitions, etc. It's a problem for scientists that how to improve their coding standard in ML model development, and who will own and maintain the code of the model."} For this issue, the practitioner agreed that it would be better if the requirements of coding standards at the ML component level were directly called out to get the corresponding resources.

Some practitioners also called out the issues encountered by ML scientists on the dependency of data pipelines and infrastructure. For the dependency of data pipelines, P5 shared that \textit{\normalsize "to train a new ML model, we need to backfill the data in the past few weeks to be used in training, and it requires non-trivial work and is burdensome"}. No or limited infrastructure support on some infra-level errors in the data pipeline was also called out to be challenging points (P9). For these issues, practitioners agreed that it makes sense to call out these requirements to the organization or leadership level, but in practice, very few supports were actually given to address them. The above evidence led to the following proposition:
\hfill\\ \\
\fbox{\begin{minipage}{25em}
\textbf{\normalsize PR4}: Requirements and detailed design of ML components with interfaces, alternatives, and expected performances should be created, with the participation of the owners of external and internal components such as data, and infrastructure. (V, SE2, SE3, SE4)
\end{minipage}}
\subsubsection{Contributions of the V-Model}  Due to the difference in skills and backgrounds of ML engineers, combined with issues in ML component dependency, certain aspects of ML components are missing in terms of requirements and design. The system boundary and responsibility of V-Model ensures that this information is created and reviewed by the internal and external owners.

\subsection{Data Engineering}

\subsubsection{Challenges in Literature}
The existing literature pointed out several issues on the data side, including difficulty in ensuring data quality and data understanding from domain experts, lack of tool support for monitoring and detecting data evolution~\cite{vogelsang2019requirements,giray2021software,whang2020data,gudivada2017data,nahar2023meta}. 
\subsubsection{Discussions from Practitioners}
Some of these issues were also reported by practitioners. One practitioner had experienced that engineers had to do multiple iterations to fix several data issues and finally get the correct data they needed for model training. Unless the engineer has very deep ML experience and expertise, it was very difficult for the engineer to get it right in the first iteration given the complexity of these data issues. One practitioner stated in their ChatGPT-based project that the quality of data (team's wiki documents) was so low that the ChatGPT output using the low-quality data turned out to work poorly (P3). The data quality may be improved by asking data engineers to do some data cleaning, but the best way is to get help from domain experts to manually edit the data, which requires much more effort. One practitioner also mentioned the model dataset is very important and it's better to make an investment to automate them rather than do much manual work to generate these datasets (P10). Another practitioner mentioned the difficulty of data labeling that was done manually via human evaluation (P8). The practitioners agreed that data as a component is a pain point that needs to be emphasized with thorough requirements, validation, and monitoring. This evidence led to the following proposition:
\hfill\\ \\
\fbox{\begin{minipage}{25em}
\textbf{\normalsize PR5}: Besides ML components and non-ML components, data should be a separate component with standalone requirements (data requirements), design synthesis, and system validation (data validation and data monitoring). Optionally, the same applies to infra as a separate component. (SE2, SE3, SE4, SE5, SE9)
\end{minipage}}
\subsubsection{Contributions of the V-Model}
Despite the complexity and iterative nature of data or infra issues, the clear boundary and responsibility of the V-Model between systems, subsystems, and components ensure the enforcement of requirements and design for data or infra components. The risk management (SE9) together with V\&V (SE5) also plays a role in mitigation when data or infra components have uncertainty.

 \subsection{Quality Assurance}

\subsubsection{Challenges in Literature}
In the existing literature, ML model teams assumed no responsibility for testing the product beyond creating a model from an accuracy perspective, while product teams in several organizations also did not have planned testing at the system level~\cite{nahar2023meta,nahar2022collaboration,liu2020emerging}. 
\subsubsection{Discussions from Practitioners}
Practitioners generally agreed on the importance and the need for testing at the system level using methods such as online monitoring metrics, load testing, and A/B testing. Some practitioners had experience handling a ticket hard to investigate: \textit{\normalsize "it's not clear if the root cause of the ticket is in the ML component or non-ML component."} (P1,P2). Practitioners agreed that it would help a lot if there were metrics, monitoring, or tools on ML and non-ML components to distinguish whether the issue is from the ML model or not.

One practitioner had multiple experiences with identifying noisy metrics during A/B testing, an issue difficult to address from the experiment platform infra (P5). Another difficulty in testing reported by practitioners is on multiple iterations of A/B testing (P1, P8). This is due to several reasons: first, the system is too complex, and the earlier assumptions on performance and business metrics are wrong. The black box of the ML model is another reason for causing multiple iterations of tuning parameters. A third reason is the gap between offline model evaluation and online testing reported by multiple practitioners (P1, P5, P6, P9). To address this, practitioners stated that after the model is launched, they would usually add a holdout experiment (P5), or add metrics to do some continuous monitoring (P8), or so-called \textit{\normalsize "acceptance test"} to check if there is any gap between online tests in the product and offline tests in the ML model (P6, P9). One practitioner observed unclear responsibility for system-level testing between ML scientists and software engineers (P6). The above evidence led to the following proposition:
\hfill\\ \\
\fbox{\begin{minipage}{25em}
\textbf{\normalsize PR6}: Verification \& Validation (V\&V) at both system, subsystem level, and component level (ML, non-ML, data, and infra), such as testing and monitoring metrics, should be enforced with identified owners. (V, SE5)
\end{minipage}}
\subsubsection{Contributions of the V-Model} V\&V (SE5) of V-Model enforce the consistency at system level, subsystem level, and component level. This consistency at the system and component level becomes important in discovering potential issues for quality assurance when different components (such as ML, data, and infra) are more unreliable when building ML-enabled systems. 

\subsection{Process}

\subsubsection{Challenges in Literature}
From a software process perspective, in the literature of ML-enabled systems, several studies have reported a lack of a good, well-defined, system-wide focused process~\cite{nahar2022collaboration,nahar2023meta,arpteg2018software,haakman2021ai}. In addition, some studies reported that the planning and estimate are difficult due to the uncertain and experimental nature of ML engineering~\cite{nahar2023meta,arpteg2018software,golendukhina2022software}. 

\subsubsection{Discussions from Practitioners}All of the practitioners agreed on the ad-hoc process for the development process of complex ML-enabled systems due to the uncertainty of science-like experiments of ML development. Most of the practitioners agreed that the concept of decomposition of the system, subsystem, and components of the V-Model helps define the responsibility boundary much more easily. This also helps explicitly call out phases that may be neglected such as requirements, V\&V, and retirement at each level. One practitioner mentioned that it's sometimes easy to forget about the system level and component level due to the complexity of ML-based projects (P2). For system Retirement / Cleanup, one practitioner had the experience of multiple A/B testing codes not being cleaned up, and this caused lots of confusion and even bugs in the system (P1). The above evidence led to the following proposition:
\hfill\\ \\
\fbox{\begin{minipage}{25em}
\textbf{\normalsize PR7}: The software development lifecycle (SDLM) for ML-enabled systems should follow layered decomposition of systems, to subsystem, to components (ML, non-ML, data, infra), with continuous in-process V\&V at system, subsystem, and component levels, continuous risk management for the uncertainty of ML, and responsibility boundaries for roles including software engineers, ML scientists, product managers, and data engineers, etc. (V, SE5)

\end{minipage}}
\subsubsection{Contributions of the V-Model} Similarly, clear system boundary and V\&V (SE5) of V-Model enforce the consistency at system level, subsystem level, and component level for the software development process in building ML-enabled systems.

 \subsection{Organization, Teams, and Responsibility}

\subsubsection{Challenges in Literature}

There are concerns in the literature about interdisciplinary collaboration and communications for roles like ML scientists and software engineers~\cite{nahar2023meta,nahar2022collaboration}. Several studies reported a lack of diverse skill sets for building ML-enabled systems, including expertise from engineering, statistics, business, UX design, operations, etc~\cite{nahar2023meta,nahar2022collaboration,john2020ai,serban2022adapting}. Some studies pointed out that poor documentation is one of the reasons for such collaboration challenges since various decisions in ML-enabled systems cannot be understood if not documented~\cite{nahar2023meta,haakman2021ai,makinen2021needs}. Other works found additional complexities in the documentation for ML engineering~\cite{nahar2023meta,chang2022understanding}. 
\subsubsection{Discussions from Practitioners}
 One practitioner raised the knowledge gap between software engineers and ML scientists such as \textit{\normalsize "pre-training, fine-tuning, model cadence, etc."} (P6) These are the concepts that software engineers need to learn. Meanwhile, the practitioner also called out the need to improve the coding quality of ML scientists in general: \textit{\normalsize "ML scientists usually develop many new models and launch them, and then find new things, but for the old models, they don't care much and stop maintaining them anymore. Our ratio of SDEs and ML scientists is 1:4, so SDEs do not have much bandwidth to help scientists. Instead, they wrote some documents about coding standards, and let scientists write the code and SDEs would review them. SDEs also created coding packages to make scientists easier to write code themselves. There were a lot of discussions within the team on this."} (P6) The practitioner agreed that documentation is, in general, a good way to address this issue. For example, it will help if ML scientists write documentation for all roles at the system level rather than for ML scientists themselves. Conversely, it also helps if software engineers write documentation about coding standards or create coding packages for ML scientists. The above evidence led to the following proposition:
\hfill\\ \\
\fbox{\begin{minipage}{25em}
\textbf{\normalsize PR8}: Documentation at the system, subsystem, and component (ML, non-ML, data, infra) levels should be created, approved by the corresponding owners and tracked for any changes, with consolidated terminology understood by all roles involved in building ML-enabled systems. (SE5, SE10)
\end{minipage}}
\subsubsection{Contributions of the V-Model} Likewise, clear system boundary, system responsibility, and V\&V (SE5) of V-Model ensures consistent and inclusive documentation that can be accessed and understood by roles with very different skills (ML scientists, software engineers, etc.) and different system level for building ML-enabled systems.

\section{Conclusions, Limitations, and Future Work}
In this work, we take a Systems Engineering lens to close the gap in collaboration challenges when building ML-enabled systems through the application of the V-Model. With the V-Model and the interviews with industrial practitioners, we collected empirical evidence and then developed 8 propositions to address the research question of \textit{\normalsize how can software process life cycles be established to address the interdisciplinary collaboration challenges from ML-enabled systems?} Our study revealed that, despite requiring additional efforts, the characteristics of V-Model, such as the system decomposition, clear boundaries and responsibilities between systems and components, and consistency of V\&V, align effectively with several collaboration challenges encountered by practitioners when building ML-enabled systems. This work makes contributions by 1) conducting one of the first exploratory studies to apply V-Model in building ML-enabled systems, and 2) addressing the gap of these collaboration challenges with a set of 8 propositions and their connections to V-Model. We recommend future research to investigate new process models, frameworks and tools that leverage these characteristics of V-Model for building ML-enabled systems.

On the other hand, limitations do exist for the use of V-Model. They include jumping between phases (e.g., from design back to requirements), failure to capture cyclical/iterative ways of software development, heavy documentation, lack of "Speed to market" and "Requirement Variability" due to its limited flexibility, agility, and adaptability to changes. These limitations could be one of the reasons that V-Model, to our knowledge, has not been adopted by organizations in practice for ML development. Thus, one future direction is to explore various ways to understand and address the trade-off between the amount of additional effort required and the actual benefits (or Return on Investment) of V-Model in building ML-enabled systems.

\begin{acks}
I gratefully thank the 11 participants for conducting the interviews, and thank the anonymous reviewers for their invaluable feedback. I extend my gratitude to CAIN for providing a platform to research works with a systems and life cycle perspective. Finally, this work would not be possible without the following people: my wife Monica Ma for her support, and my parents Baofu Wu and Jing Wang for their support, especially for helping take care of my child when needed. 
\end{acks}

\normalfont
\bibliographystyle{ACM-Reference-Format}
\bibliography{sample-base}


\begin{thebibliography}{77}


\ifx \showCODEN    \undefined \def \showCODEN     #1{\unskip}     \fi
\ifx \showDOI      \undefined \def \showDOI       #1{#1}\fi
\ifx \showISBNx    \undefined \def \showISBNx     #1{\unskip}     \fi
\ifx \showISBNxiii \undefined \def \showISBNxiii  #1{\unskip}     \fi
\ifx \showISSN     \undefined \def \showISSN      #1{\unskip}     \fi
\ifx \showLCCN     \undefined \def \showLCCN      #1{\unskip}     \fi
\ifx \shownote     \undefined \def \shownote      #1{#1}          \fi
\ifx \showarticletitle \undefined \def \showarticletitle #1{#1}   \fi
\ifx \showURL      \undefined \def \showURL       {\relax}        \fi
\providecommand\bibfield[2]{#2}
\providecommand\bibinfo[2]{#2}
\providecommand\natexlab[1]{#1}
\providecommand\showeprint[2][]{arXiv:#2}

\bibitem[Akkiraju et~al\mbox{.}(2020)]%
        {akkiraju2020characterizing}
\bibfield{author}{\bibinfo{person}{Rama Akkiraju}, \bibinfo{person}{Vibha
  Sinha}, \bibinfo{person}{Anbang Xu}, \bibinfo{person}{Jalal Mahmud},
  \bibinfo{person}{Pritam Gundecha}, \bibinfo{person}{Zhe Liu},
  \bibinfo{person}{Xiaotong Liu}, {and} \bibinfo{person}{John Schumacher}.}
  \bibinfo{year}{2020}\natexlab{}.
\newblock \showarticletitle{Characterizing machine learning processes: A
  maturity framework}. In \bibinfo{booktitle}{\emph{Business Process
  Management: 18th International Conference, BPM 2020, Seville, Spain,
  September 13--18, 2020, Proceedings 18}}. Springer, \bibinfo{pages}{17--31}.
\newblock


\bibitem[Amershi et~al\mbox{.}(2019)]%
        {amershi2019software}
\bibfield{author}{\bibinfo{person}{Saleema Amershi}, \bibinfo{person}{Andrew
  Begel}, \bibinfo{person}{Christian Bird}, \bibinfo{person}{Robert DeLine},
  \bibinfo{person}{Harald Gall}, \bibinfo{person}{Ece Kamar},
  \bibinfo{person}{Nachiappan Nagappan}, \bibinfo{person}{Besmira Nushi}, {and}
  \bibinfo{person}{Thomas Zimmermann}.} \bibinfo{year}{2019}\natexlab{}.
\newblock \showarticletitle{Software engineering for machine learning: A case
  study}. In \bibinfo{booktitle}{\emph{2019 IEEE/ACM 41st International
  Conference on Software Engineering: Software Engineering in Practice
  (ICSE-SEIP)}}. IEEE, \bibinfo{pages}{291--300}.
\newblock


\bibitem[Arpteg et~al\mbox{.}(2018)]%
        {arpteg2018software}
\bibfield{author}{\bibinfo{person}{Anders Arpteg}, \bibinfo{person}{Bj{\"o}rn
  Brinne}, \bibinfo{person}{Luka Crnkovic-Friis}, {and} \bibinfo{person}{Jan
  Bosch}.} \bibinfo{year}{2018}\natexlab{}.
\newblock \showarticletitle{Software engineering challenges of deep learning}.
  In \bibinfo{booktitle}{\emph{2018 44th euromicro conference on software
  engineering and advanced applications (SEAA)}}. IEEE,
  \bibinfo{pages}{50--59}.
\newblock


\bibitem[Balaji and Murugaiyan(2012)]%
        {balaji2012waterfall}
\bibfield{author}{\bibinfo{person}{Sundramoorthy Balaji} {and}
  \bibinfo{person}{M~Sundararajan Murugaiyan}.}
  \bibinfo{year}{2012}\natexlab{}.
\newblock \showarticletitle{Waterfall vs. V-Model vs. Agile: A comparative
  study on SDLC}.
\newblock \bibinfo{journal}{\emph{International Journal of Information
  Technology and Business Management}} \bibinfo{volume}{2}, \bibinfo{number}{1}
  (\bibinfo{year}{2012}), \bibinfo{pages}{26--30}.
\newblock


\bibitem[Beale and Bonometti(2006)]%
        {beale2006systems}
\bibfield{author}{\bibinfo{person}{David Beale} {and} \bibinfo{person}{Joseph
  Bonometti}.} \bibinfo{year}{2006}\natexlab{}.
\newblock \showarticletitle{Systems engineering (SE)-the systems design
  process}.
\newblock \bibinfo{journal}{\emph{The Lunar Engineering Handbook, Auburg
  University, Auburn}} (\bibinfo{year}{2006}).
\newblock


\bibitem[Bersson et~al\mbox{.}(2012)]%
        {bersson2012framework}
\bibfield{author}{\bibinfo{person}{Thomas~F Bersson}, \bibinfo{person}{Thomas
  Mazzuchi}, {and} \bibinfo{person}{Shahram Sarkani}.}
  \bibinfo{year}{2012}\natexlab{}.
\newblock \showarticletitle{A framework for application of system engineering
  process models to sustainable design of high performance buildings}.
\newblock \bibinfo{journal}{\emph{Journal of Green Building}}
  \bibinfo{volume}{7}, \bibinfo{number}{3} (\bibinfo{year}{2012}),
  \bibinfo{pages}{171--192}.
\newblock


\bibitem[Bitektine(2008)]%
        {bitektine2008prospective}
\bibfield{author}{\bibinfo{person}{Alex Bitektine}.}
  \bibinfo{year}{2008}\natexlab{}.
\newblock \showarticletitle{Prospective case study design: Qualitative method
  for deductive theory testing}.
\newblock \bibinfo{journal}{\emph{Organizational research methods}}
  \bibinfo{volume}{11}, \bibinfo{number}{1} (\bibinfo{year}{2008}),
  \bibinfo{pages}{160--180}.
\newblock


\bibitem[Blanchard and Fabrycky(2011)]%
        {Blanchard-Fabrycky-SystemsEngineering}
\bibfield{author}{\bibinfo{person}{Benjamin~S. Blanchard} {and}
  \bibinfo{person}{Wolter~J. Fabrycky}.} \bibinfo{year}{2011}\natexlab{}.
\newblock \bibinfo{booktitle}{\emph{Systems Engineering and Analysis}}.
\newblock \bibinfo{publisher}{Prentice Hall}.
\newblock


\bibitem[Bodner and Rouse(2009)]%
        {bodner2009handbook}
\bibfield{author}{\bibinfo{person}{DA Bodner} {and} \bibinfo{person}{WB
  Rouse}.} \bibinfo{year}{2009}\natexlab{}.
\newblock \showarticletitle{Handbook of systems engineering and management}.
\newblock \bibinfo{journal}{\emph{Wiley. chapter Organizational Simulation}}
  (\bibinfo{year}{2009}).
\newblock


\bibitem[Borg et~al\mbox{.}(2019)]%
        {borg2019safely}
\bibfield{author}{\bibinfo{person}{Markus Borg}, \bibinfo{person}{Cristofer
  Englund}, \bibinfo{person}{Krzysztof Wnuk}, \bibinfo{person}{Boris Duran},
  \bibinfo{person}{Christoffer Levandowski}, \bibinfo{person}{Shenjian Gao},
  \bibinfo{person}{Yanwen Tan}, \bibinfo{person}{Henrik Kaijser},
  \bibinfo{person}{Henrik L\"onn}, {and} \bibinfo{person}{Jonas T\"ornqvist}.}
  \bibinfo{year}{2019}\natexlab{}.
\newblock \showarticletitle{Safely entering the deep: A review of verification
  and validation for machine learning and a challenge elicitation in the
  automotive industry}.
\newblock \bibinfo{journal}{\emph{Journal of Automotive Software Engineering}}
  \bibinfo{volume}{1}, \bibinfo{number}{1} (\bibinfo{year}{2019}),
  \bibinfo{pages}{1--19}.
\newblock


\bibitem[Bosch et~al\mbox{.}(2021)]%
        {bosch2021engineering}
\bibfield{author}{\bibinfo{person}{Jan Bosch},
  \bibinfo{person}{Helena~Holmstr{\"o}m Olsson}, {and} \bibinfo{person}{Ivica
  Crnkovic}.} \bibinfo{year}{2021}\natexlab{}.
\newblock \showarticletitle{Engineering ai systems: A research agenda}.
\newblock \bibinfo{journal}{\emph{Artificial Intelligence Paradigms for Smart
  Cyber-Physical Systems}} (\bibinfo{year}{2021}), \bibinfo{pages}{1--19}.
\newblock


\bibitem[Braiek and Khomh(2020)]%
        {braiek2020testing}
\bibfield{author}{\bibinfo{person}{Houssem~Ben Braiek} {and}
  \bibinfo{person}{Foutse Khomh}.} \bibinfo{year}{2020}\natexlab{}.
\newblock \showarticletitle{On testing machine learning programs}.
\newblock \bibinfo{journal}{\emph{Journal of Systems and Software}}
  \bibinfo{volume}{164} (\bibinfo{year}{2020}), \bibinfo{pages}{110542}.
\newblock


\bibitem[Braude and Bernstein(2016)]%
        {braude2016software}
\bibfield{author}{\bibinfo{person}{Eric~J Braude} {and}
  \bibinfo{person}{Michael~E Bernstein}.} \bibinfo{year}{2016}\natexlab{}.
\newblock \bibinfo{booktitle}{\emph{Software engineering: modern approaches}}.
\newblock \bibinfo{publisher}{Waveland Press}.
\newblock


\bibitem[Br{\"o}hl(1993)]%
        {brohl1993v}
\bibfield{author}{\bibinfo{person}{Adolf-Peter Br{\"o}hl}.}
  \bibinfo{year}{1993}\natexlab{}.
\newblock \bibinfo{booktitle}{\emph{Das V-Modell: Der Standard f{\"u}r die
  Softwareentwicklung mit Praxisleitfaden}}.
\newblock \bibinfo{publisher}{Oldenbourg}.
\newblock


\bibitem[Cechini et~al\mbox{.}(2009)]%
        {cechini2009systems}
\bibfield{author}{\bibinfo{person}{F Cechini}, \bibinfo{person}{R Ice}, {and}
  \bibinfo{person}{D Binkley}.} \bibinfo{year}{2009}\natexlab{}.
\newblock \showarticletitle{Systems Engineering Guidebook for Intelligent
  Transportation Systems}.
\newblock \bibinfo{journal}{\emph{California Division of the United States
  Department of Transportation Federal Highway Administration and the
  California Department of Transportation}} (\bibinfo{year}{2009}).
\newblock


\bibitem[Cederbladh et~al\mbox{.}(2023)]%
        {cederbladh2023early}
\bibfield{author}{\bibinfo{person}{Johan Cederbladh}, \bibinfo{person}{Antonio
  Cicchetti}, {and} \bibinfo{person}{Jagadish Suryadevara}.}
  \bibinfo{year}{2023}\natexlab{}.
\newblock \showarticletitle{Early Validation and Verification of System
  Behaviour in Model-Based Systems Engineering: A Systematic Literature
  Review}.
\newblock \bibinfo{journal}{\emph{ACM Transactions on Software Engineering and
  Methodology}} (\bibinfo{year}{2023}).
\newblock


\bibitem[Chang and Custis(2022)]%
        {chang2022understanding}
\bibfield{author}{\bibinfo{person}{Jiyoo Chang} {and}
  \bibinfo{person}{Christine Custis}.} \bibinfo{year}{2022}\natexlab{}.
\newblock \showarticletitle{Understanding Implementation Challenges in Machine
  Learning Documentation}.
\newblock In \bibinfo{booktitle}{\emph{Equity and Access in Algorithms,
  Mechanisms, and Optimization}}. \bibinfo{pages}{1--8}.
\newblock


\bibitem[Dove et~al\mbox{.}(2017)]%
        {dove2017ux}
\bibfield{author}{\bibinfo{person}{Graham Dove}, \bibinfo{person}{Kim Halskov},
  \bibinfo{person}{Jodi Forlizzi}, {and} \bibinfo{person}{John Zimmerman}.}
  \bibinfo{year}{2017}\natexlab{}.
\newblock \showarticletitle{UX design innovation: Challenges for working with
  machine learning as a design material}. In
  \bibinfo{booktitle}{\emph{Proceedings of the 2017 chi conference on human
  factors in computing systems}}. \bibinfo{pages}{278--288}.
\newblock


\bibitem[Ericson et~al\mbox{.}(2017)]%
        {ericson2017team}
\bibfield{author}{\bibinfo{person}{Gary Ericson},
  \bibinfo{person}{William~Anton Rohm}, \bibinfo{person}{Jos{\'e}e Martens},
  \bibinfo{person}{Kent Sharkey}, \bibinfo{person}{Craig Casey},
  \bibinfo{person}{Beth Harvey}, {and} \bibinfo{person}{Nick Schonning}.}
  \bibinfo{year}{2017}\natexlab{}.
\newblock \showarticletitle{Team data science process documentation}.
\newblock \bibinfo{journal}{\emph{Retrieved April}}  \bibinfo{volume}{11}
  (\bibinfo{year}{2017}), \bibinfo{pages}{2019}.
\newblock


\bibitem[Fanson(2010)]%
        {fanson2010lessons}
\bibfield{author}{\bibinfo{person}{James Fanson}.}
  \bibinfo{year}{2010}\natexlab{}.
\newblock \showarticletitle{Lessons learned from the Kepler Mission and space
  telescope management}. In \bibinfo{booktitle}{\emph{An Optical Believe It or
  Not: Key Lessons Learned II}}, Vol.~\bibinfo{volume}{7796}. SPIE,
  \bibinfo{pages}{25--30}.
\newblock


\bibitem[Georgiadis et~al\mbox{.}(2013)]%
        {georgiadis2013using}
\bibfield{author}{\bibinfo{person}{Daniel~R Georgiadis},
  \bibinfo{person}{Thomas~A Mazzuchi}, {and} \bibinfo{person}{Shahram
  Sarkani}.} \bibinfo{year}{2013}\natexlab{}.
\newblock \showarticletitle{Using multi criteria decision making in analysis of
  alternatives for selection of enabling technology}.
\newblock \bibinfo{journal}{\emph{Systems Engineering}} \bibinfo{volume}{16},
  \bibinfo{number}{3} (\bibinfo{year}{2013}), \bibinfo{pages}{287--303}.
\newblock


\bibitem[Giray(2021)]%
        {giray2021software}
\bibfield{author}{\bibinfo{person}{G{\"o}rkem Giray}.}
  \bibinfo{year}{2021}\natexlab{}.
\newblock \showarticletitle{A software engineering perspective on engineering
  machine learning systems: State of the art and challenges}.
\newblock \bibinfo{journal}{\emph{Journal of Systems and Software}}
  \bibinfo{volume}{180} (\bibinfo{year}{2021}), \bibinfo{pages}{111031}.
\newblock


\bibitem[Golendukhina et~al\mbox{.}(2022)]%
        {golendukhina2022software}
\bibfield{author}{\bibinfo{person}{Valentina Golendukhina},
  \bibinfo{person}{Valentina Lenarduzzi}, {and} \bibinfo{person}{Michael
  Felderer}.} \bibinfo{year}{2022}\natexlab{}.
\newblock \showarticletitle{What is software quality for AI engineers? Towards
  a thinning of the fog}. In \bibinfo{booktitle}{\emph{Proceedings of the 1st
  International Conference on AI Engineering: Software Engineering for AI}}.
  \bibinfo{pages}{1--9}.
\newblock


\bibitem[Graessler and Hentze(2020)]%
        {graessler2020new}
\bibfield{author}{\bibinfo{person}{Iris Graessler} {and}
  \bibinfo{person}{Julian Hentze}.} \bibinfo{year}{2020}\natexlab{}.
\newblock \showarticletitle{The new V-Model of VDI 2206 and its validation}.
\newblock \bibinfo{journal}{\emph{at-Automatisierungstechnik}}
  \bibinfo{volume}{68}, \bibinfo{number}{5} (\bibinfo{year}{2020}),
  \bibinfo{pages}{312--324}.
\newblock


\bibitem[Gudivada et~al\mbox{.}(2017)]%
        {gudivada2017data}
\bibfield{author}{\bibinfo{person}{Venkat Gudivada}, \bibinfo{person}{Amy
  Apon}, {and} \bibinfo{person}{Junhua Ding}.} \bibinfo{year}{2017}\natexlab{}.
\newblock \showarticletitle{Data quality considerations for big data and
  machine learning: Going beyond data cleaning and transformations}.
\newblock \bibinfo{journal}{\emph{International Journal on Advances in
  Software}} \bibinfo{volume}{10}, \bibinfo{number}{1} (\bibinfo{year}{2017}),
  \bibinfo{pages}{1--20}.
\newblock


\bibitem[Haakman et~al\mbox{.}(2021)]%
        {haakman2021ai}
\bibfield{author}{\bibinfo{person}{Mark Haakman}, \bibinfo{person}{Lu{\'\i}s
  Cruz}, \bibinfo{person}{Hennie Huijgens}, {and} \bibinfo{person}{Arie van
  Deursen}.} \bibinfo{year}{2021}\natexlab{}.
\newblock \showarticletitle{AI lifecycle models need to be revised: An
  exploratory study in Fintech}.
\newblock \bibinfo{journal}{\emph{Empirical Software Engineering}}
  \bibinfo{volume}{26} (\bibinfo{year}{2021}), \bibinfo{pages}{1--29}.
\newblock


\bibitem[Haberfellner et~al\mbox{.}(2019)]%
        {haberfellner2019systems}
\bibfield{author}{\bibinfo{person}{Reinhard Haberfellner},
  \bibinfo{person}{Olivier De~Weck}, \bibinfo{person}{Ernst Fricke}, {and}
  \bibinfo{person}{Siegfried V{\"o}ssner}.} \bibinfo{year}{2019}\natexlab{}.
\newblock \bibinfo{booktitle}{\emph{Systems engineering: fundamentals and
  applications}}.
\newblock \bibinfo{publisher}{Springer}.
\newblock


\bibitem[Huang et~al\mbox{.}(2011)]%
        {huang2011adversarial}
\bibfield{author}{\bibinfo{person}{Ling Huang}, \bibinfo{person}{Anthony~D.
  Joseph}, \bibinfo{person}{Blaine Nelson}, \bibinfo{person}{Benjamin~IP
  Rubinstein}, {and} \bibinfo{person}{J.~Doug Tygar}.}
  \bibinfo{year}{2011}\natexlab{}.
\newblock \showarticletitle{Adversarial machine learning}. In
  \bibinfo{booktitle}{\emph{Proceedings of the 4th ACM Workshop on Security and
  Artificial Intelligence}}. \bibinfo{pages}{43--58}.
\newblock


\bibitem[Huang et~al\mbox{.}(2020)]%
        {huang2020survey}
\bibfield{author}{\bibinfo{person}{Xiaowei Huang}, \bibinfo{person}{Daniel
  Kroening}, \bibinfo{person}{Wenjie Ruan}, \bibinfo{person}{James Sharp},
  \bibinfo{person}{Youcheng Sun}, \bibinfo{person}{Emese Thamo},
  \bibinfo{person}{Min Wu}, {and} \bibinfo{person}{Xinping Yi}.}
  \bibinfo{year}{2020}\natexlab{}.
\newblock \showarticletitle{A survey of safety and trustworthiness of deep
  neural networks: Verification, testing, adversarial attack and defence, and
  interpretability}.
\newblock \bibinfo{journal}{\emph{Computer Science Review}}
  \bibinfo{volume}{37} (\bibinfo{year}{2020}), \bibinfo{pages}{100270}.
\newblock


\bibitem[John et~al\mbox{.}(2020)]%
        {john2020ai}
\bibfield{author}{\bibinfo{person}{Meenu~Mary John},
  \bibinfo{person}{Helena~Holmstr{\"o}m Olsson}, {and} \bibinfo{person}{Jan
  Bosch}.} \bibinfo{year}{2020}\natexlab{}.
\newblock \showarticletitle{Ai deployment architecture: Multi-case study for
  key factor identification}. In \bibinfo{booktitle}{\emph{2020 27th
  Asia-Pacific Software Engineering Conference (APSEC)}}. IEEE,
  \bibinfo{pages}{395--404}.
\newblock


\bibitem[Kim et~al\mbox{.}(2017)]%
        {kim2017data}
\bibfield{author}{\bibinfo{person}{Miryung Kim}, \bibinfo{person}{Thomas
  Zimmermann}, \bibinfo{person}{Robert DeLine}, {and} \bibinfo{person}{Andrew
  Begel}.} \bibinfo{year}{2017}\natexlab{}.
\newblock \showarticletitle{Data scientists in software teams: State of the art
  and challenges}.
\newblock \bibinfo{journal}{\emph{IEEE Transactions on Software Engineering}}
  \bibinfo{volume}{44}, \bibinfo{number}{11} (\bibinfo{year}{2017}),
  \bibinfo{pages}{1024--1038}.
\newblock


\bibitem[Kossiakoff et~al\mbox{.}(2011)]%
        {Kossiakoff-SystemsEngineeringPrinciplesAndPractice}
\bibfield{author}{\bibinfo{person}{Alexander Kossiakoff},
  \bibinfo{person}{William~N. Sweet}, \bibinfo{person}{Sam Seymour}, {and}
  \bibinfo{person}{Steven~M. Biemer}.} \bibinfo{year}{2011}\natexlab{}.
\newblock \bibinfo{booktitle}{\emph{Systems Engineering: Principles and
  Practice}}.
\newblock \bibinfo{publisher}{John Wiley \& Sons}.
\newblock


\bibitem[Laato et~al\mbox{.}(2022)]%
        {laato2022integrating}
\bibfield{author}{\bibinfo{person}{Samuli Laato}, \bibinfo{person}{Matti
  M{\"a}ntym{\"a}ki}, \bibinfo{person}{Matti Minkkinen}, \bibinfo{person}{Teemu
  Birkstedt}, \bibinfo{person}{AKM Islam}, {and} \bibinfo{person}{Denis
  Dennehy}.} \bibinfo{year}{2022}\natexlab{}.
\newblock \showarticletitle{Integrating machine learning with software
  development lifecycles: Insights from experts}.
\newblock  (\bibinfo{year}{2022}).
\newblock


\bibitem[Lewis et~al\mbox{.}(2021)]%
        {lewis2021software}
\bibfield{author}{\bibinfo{person}{Grace~A Lewis}, \bibinfo{person}{Ipek
  Ozkaya}, {and} \bibinfo{person}{Xiwei Xu}.} \bibinfo{year}{2021}\natexlab{}.
\newblock \showarticletitle{Software architecture challenges for ml systems}.
  In \bibinfo{booktitle}{\emph{2021 IEEE International Conference on Software
  Maintenance and Evolution (ICSME)}}. IEEE, \bibinfo{pages}{634--638}.
\newblock


\bibitem[Li et~al\mbox{.}(2022)]%
        {li2022testing}
\bibfield{author}{\bibinfo{person}{Shuyue Li}, \bibinfo{person}{Jiaqi Guo},
  \bibinfo{person}{Jian-Guang Lou}, \bibinfo{person}{Ming Fan},
  \bibinfo{person}{Ting Liu}, {and} \bibinfo{person}{Dongmei Zhang}.}
  \bibinfo{year}{2022}\natexlab{}.
\newblock \showarticletitle{Testing machine learning systems in industry: an
  empirical study}. In \bibinfo{booktitle}{\emph{Proceedings of the 44th
  International Conference on Software Engineering: Software Engineering in
  Practice}}. \bibinfo{pages}{263--272}.
\newblock


\bibitem[Liu et~al\mbox{.}(2020)]%
        {liu2020emerging}
\bibfield{author}{\bibinfo{person}{Hanyan Liu}, \bibinfo{person}{Samuel Eksmo},
  \bibinfo{person}{Johan Risberg}, {and} \bibinfo{person}{Regina Hebig}.}
  \bibinfo{year}{2020}\natexlab{}.
\newblock \showarticletitle{Emerging and changing tasks in the development
  process for machine learning systems}. In
  \bibinfo{booktitle}{\emph{Proceedings of the international conference on
  software and system processes}}. \bibinfo{pages}{125--134}.
\newblock


\bibitem[Liu et~al\mbox{.}(2018)]%
        {liu2018survey}
\bibfield{author}{\bibinfo{person}{Qiang Liu}, \bibinfo{person}{Pan Li},
  \bibinfo{person}{Wentao Zhao}, \bibinfo{person}{Wei Cai},
  \bibinfo{person}{Shui Yu}, {and} \bibinfo{person}{Victor~CM Leung}.}
  \bibinfo{year}{2018}\natexlab{}.
\newblock \showarticletitle{A survey on security threats and defensive
  techniques of machine learning: A data driven view}.
\newblock \bibinfo{journal}{\emph{IEEE access}}  \bibinfo{volume}{6}
  (\bibinfo{year}{2018}), \bibinfo{pages}{12103--12117}.
\newblock


\bibitem[Maier and Rechtin(2009)]%
        {Maier-Rechtin-ArtOfSystemsArchitecting}
\bibfield{author}{\bibinfo{person}{Mark~W. Maier} {and}
  \bibinfo{person}{Eberhardt Rechtin}.} \bibinfo{year}{2009}\natexlab{}.
\newblock \bibinfo{booktitle}{\emph{The Art of Systems Architecting}}.
\newblock \bibinfo{publisher}{CRC Press}.
\newblock


\bibitem[M{\"a}kinen et~al\mbox{.}(2021)]%
        {makinen2021needs}
\bibfield{author}{\bibinfo{person}{Sasu M{\"a}kinen}, \bibinfo{person}{Henrik
  Skogstr{\"o}m}, \bibinfo{person}{Eero Laaksonen}, {and}
  \bibinfo{person}{Tommi Mikkonen}.} \bibinfo{year}{2021}\natexlab{}.
\newblock \showarticletitle{Who needs MLOps: What data scientists seek to
  accomplish and how can MLOps help?}. In \bibinfo{booktitle}{\emph{2021
  IEEE/ACM 1st Workshop on AI Engineering-Software Engineering for AI (WAIN)}}.
  IEEE, \bibinfo{pages}{109--112}.
\newblock


\bibitem[Mart{\'\i}nez-Plumed et~al\mbox{.}(2019)]%
        {martinez2019crisp}
\bibfield{author}{\bibinfo{person}{Fernando Mart{\'\i}nez-Plumed},
  \bibinfo{person}{Lidia Contreras-Ochando}, \bibinfo{person}{Cesar Ferri},
  \bibinfo{person}{Jos{\'e} Hern{\'a}ndez-Orallo}, \bibinfo{person}{Meelis
  Kull}, \bibinfo{person}{Nicolas Lachiche},
  \bibinfo{person}{Mar{\'\i}a~Jos{\'e} Ram{\'\i}rez-Quintana}, {and}
  \bibinfo{person}{Peter Flach}.} \bibinfo{year}{2019}\natexlab{}.
\newblock \showarticletitle{CRISP-DM twenty years later: From data mining
  processes to data science trajectories}.
\newblock \bibinfo{journal}{\emph{IEEE Transactions on Knowledge and Data
  Engineering}} \bibinfo{volume}{33}, \bibinfo{number}{8}
  (\bibinfo{year}{2019}), \bibinfo{pages}{3048--3061}.
\newblock


\bibitem[Mathieson et~al\mbox{.}(2020)]%
        {mathieson2020systems}
\bibfield{author}{\bibinfo{person}{John~TJ Mathieson}, \bibinfo{person}{Thomas
  Mazzuchi}, {and} \bibinfo{person}{Shahram Sarkani}.}
  \bibinfo{year}{2020}\natexlab{}.
\newblock \showarticletitle{The systems engineering DevOps lemniscate and
  model-based system operations}.
\newblock \bibinfo{journal}{\emph{IEEE Systems Journal}} \bibinfo{volume}{15},
  \bibinfo{number}{3} (\bibinfo{year}{2020}), \bibinfo{pages}{3980--3991}.
\newblock


\bibitem[McGraw et~al\mbox{.}(2020)]%
        {mcgraw2020architectural}
\bibfield{author}{\bibinfo{person}{Gary McGraw}, \bibinfo{person}{Harold
  Figueroa}, \bibinfo{person}{Victor Shepardson}, {and} \bibinfo{person}{Richie
  Bonett}.} \bibinfo{year}{2020}\natexlab{}.
\newblock \bibinfo{booktitle}{\emph{An architectural risk analysis of machine
  learning systems: Toward more secure machine learning}}.
\newblock \bibinfo{type}{{T}echnical {R}eport}.
  \bibinfo{institution}{Berryville Institute of Machine Learning, v 1.0}.
\newblock


\bibitem[M\o{}ller et~al\mbox{.}(2020)]%
        {moller2020data}
\bibfield{author}{\bibinfo{person}{Naja~Holten M\o{}ller},
  \bibinfo{person}{Claus Bossen}, \bibinfo{person}{Kathleen~H. Pine},
  \bibinfo{person}{Trine~Rask Nielsen}, {and} \bibinfo{person}{Gina Neff}.}
  \bibinfo{year}{2020}\natexlab{}.
\newblock \showarticletitle{Who does the work of data?}
\newblock \bibinfo{journal}{\emph{Interactions}} \bibinfo{volume}{27},
  \bibinfo{number}{3} (\bibinfo{year}{2020}), \bibinfo{pages}{52--55}.
\newblock


\bibitem[Nahar et~al\mbox{.}(2023)]%
        {nahar2023meta}
\bibfield{author}{\bibinfo{person}{Nadia Nahar}, \bibinfo{person}{Haoran
  Zhang}, \bibinfo{person}{Grace Lewis}, \bibinfo{person}{Shurui Zhou}, {and}
  \bibinfo{person}{Christian K{\"a}stner}.} \bibinfo{year}{2023}\natexlab{}.
\newblock \showarticletitle{A Meta-Summary of Challenges in Building Products
  with ML Components--Collecting Experiences from 4758+ Practitioners}.
\newblock \bibinfo{journal}{\emph{arXiv preprint arXiv:2304.00078}}
  (\bibinfo{year}{2023}).
\newblock


\bibitem[Nahar et~al\mbox{.}(2022)]%
        {nahar2022collaboration}
\bibfield{author}{\bibinfo{person}{Nadia Nahar}, \bibinfo{person}{Shurui Zhou},
  \bibinfo{person}{Grace Lewis}, {and} \bibinfo{person}{Christian
  K{\"a}stner}.} \bibinfo{year}{2022}\natexlab{}.
\newblock \showarticletitle{Collaboration challenges in building ml-enabled
  systems: Communication, documentation, engineering, and process}. In
  \bibinfo{booktitle}{\emph{Proceedings of the 44th International Conference on
  Software Engineering}}. \bibinfo{pages}{413--425}.
\newblock


\bibitem[Namvar et~al\mbox{.}(2023)]%
        {namvar2023beyond}
\bibfield{author}{\bibinfo{person}{Morteza Namvar}, \bibinfo{person}{Ali
  Intezari}, \bibinfo{person}{Saeed Akhlaghpour}, {and}
  \bibinfo{person}{Justin~P Brienza}.} \bibinfo{year}{2023}\natexlab{}.
\newblock \showarticletitle{Beyond effective use: Integrating wise reasoning in
  machine learning development}.
\newblock \bibinfo{journal}{\emph{International Journal of Information
  Management}}  \bibinfo{volume}{69} (\bibinfo{year}{2023}),
  \bibinfo{pages}{102566}.
\newblock


\bibitem[Niven and Lamorte(2016)]%
        {niven2016objectives}
\bibfield{author}{\bibinfo{person}{Paul~R Niven} {and} \bibinfo{person}{Ben
  Lamorte}.} \bibinfo{year}{2016}\natexlab{}.
\newblock \bibinfo{booktitle}{\emph{Objectives and key results: Driving focus,
  alignment, and engagement with OKRs}}.
\newblock \bibinfo{publisher}{John Wiley \& Sons}.
\newblock


\bibitem[O'Leary and Uchida(2020)]%
        {o2020common}
\bibfield{author}{\bibinfo{person}{Katie O'Leary} {and} \bibinfo{person}{Makoto
  Uchida}.} \bibinfo{year}{2020}\natexlab{}.
\newblock \showarticletitle{Common problems with creating machine learning
  pipelines from existing code}.
\newblock  (\bibinfo{year}{2020}).
\newblock


\bibitem[Ozkaya(2020)]%
        {ozkaya2020really}
\bibfield{author}{\bibinfo{person}{Ipek Ozkaya}.}
  \bibinfo{year}{2020}\natexlab{}.
\newblock \showarticletitle{What is really different in engineering AI-enabled
  systems?}
\newblock \bibinfo{journal}{\emph{IEEE software}} \bibinfo{volume}{37},
  \bibinfo{number}{4} (\bibinfo{year}{2020}), \bibinfo{pages}{3--6}.
\newblock


\bibitem[Petersen et~al\mbox{.}(2022)]%
        {petersen2022towards}
\bibfield{author}{\bibinfo{person}{Patrick Petersen}, \bibinfo{person}{Hanno
  Stage}, \bibinfo{person}{Jacob Langner}, \bibinfo{person}{Lennart Ries},
  \bibinfo{person}{Philipp Rigoll}, \bibinfo{person}{Carl~Philipp Hohl}, {and}
  \bibinfo{person}{Eric Sax}.} \bibinfo{year}{2022}\natexlab{}.
\newblock \showarticletitle{Towards a Data Engineering Process in Data-Driven
  Systems Engineering}. In \bibinfo{booktitle}{\emph{2022 IEEE International
  Symposium on Systems Engineering (ISSE)}}. IEEE, \bibinfo{pages}{1--8}.
\newblock


\bibitem[Ranawana and Karunananda(2021)]%
        {ranawana2021agile}
\bibfield{author}{\bibinfo{person}{Romesh Ranawana} {and}
  \bibinfo{person}{Asoka~S Karunananda}.} \bibinfo{year}{2021}\natexlab{}.
\newblock \showarticletitle{An agile software development life cycle model for
  machine learning application development}. In \bibinfo{booktitle}{\emph{2021
  5th SLAAI International Conference on Artificial Intelligence (SLAAI-ICAI)}}.
  IEEE, \bibinfo{pages}{1--6}.
\newblock


\bibitem[Riccio et~al\mbox{.}(2020)]%
        {riccio2020testing}
\bibfield{author}{\bibinfo{person}{Vincenzo Riccio}, \bibinfo{person}{Gunel
  Jahangirova}, \bibinfo{person}{Andrea Stocco}, \bibinfo{person}{Nargiz
  Humbatova}, \bibinfo{person}{Michael Weiss}, {and} \bibinfo{person}{Paolo
  Tonella}.} \bibinfo{year}{2020}\natexlab{}.
\newblock \showarticletitle{Testing machine learning based systems: a
  systematic mapping}.
\newblock \bibinfo{journal}{\emph{Empirical Software Engineering}}
  (\bibinfo{year}{2020}), \bibinfo{pages}{1--62}.
\newblock


\bibitem[Ritchie and Spencer(2002)]%
        {ritchie2002qualitative}
\bibfield{author}{\bibinfo{person}{Jane Ritchie} {and} \bibinfo{person}{Liz
  Spencer}.} \bibinfo{year}{2002}\natexlab{}.
\newblock \showarticletitle{Qualitative data analysis for applied policy
  research}.
\newblock In \bibinfo{booktitle}{\emph{Analyzing qualitative data}}.
  \bibinfo{publisher}{Routledge}, \bibinfo{pages}{173--194}.
\newblock


\bibitem[Sage and Rouse(2014)]%
        {sage2014handbook}
\bibfield{author}{\bibinfo{person}{Andrew~P Sage} {and}
  \bibinfo{person}{William~B Rouse}.} \bibinfo{year}{2014}\natexlab{}.
\newblock \bibinfo{booktitle}{\emph{Handbook of systems engineering and
  management}}.
\newblock \bibinfo{publisher}{John Wiley \& Sons}.
\newblock


\bibitem[Salay and Czarnecki(2018)]%
        {salay2018using}
\bibfield{author}{\bibinfo{person}{Rick Salay} {and} \bibinfo{person}{Krzysztof
  Czarnecki}.} \bibinfo{year}{2018}\natexlab{}.
\newblock \showarticletitle{Using machine learning safely in automotive
  software: An assessment and adaption of software process requirements in ISO
  26262}.
\newblock \bibinfo{journal}{\emph{arXiv preprint arXiv:1808.01614}}
  (\bibinfo{year}{2018}).
\newblock


\bibitem[Salay et~al\mbox{.}(2017)]%
        {salay2017analysis}
\bibfield{author}{\bibinfo{person}{Rick Salay}, \bibinfo{person}{Rodrigo
  Queiroz}, {and} \bibinfo{person}{Krzysztof Czarnecki}.}
  \bibinfo{year}{2017}\natexlab{}.
\newblock \showarticletitle{An analysis of ISO 26262: Using machine learning
  safely in automotive software}.
\newblock \bibinfo{journal}{\emph{arXiv preprint arXiv:1709.02435}}
  (\bibinfo{year}{2017}).
\newblock


\bibitem[Sambasivan et~al\mbox{.}(2021)]%
        {sambasivan2021data}
\bibfield{author}{\bibinfo{person}{Nithya Sambasivan}, \bibinfo{person}{Shivani
  Kapania}, \bibinfo{person}{Hannah Highfill}, \bibinfo{person}{Diana Akrong},
  \bibinfo{person}{Praveen Paritosh}, {and} \bibinfo{person}{Lora~M. Aroyo}.}
  \bibinfo{year}{2021}\natexlab{}.
\newblock \showarticletitle{“Everyone wants to do the model work, not the
  data work”: Data Cascades in High-Stakes AI}. In
  \bibinfo{booktitle}{\emph{Proceedings of the Conference on Human Factors in
  Computing Systems}}. \bibinfo{pages}{1--15}.
\newblock


\bibitem[Sambasivan and Veeraraghavan(2022)]%
        {sambasivan2022deskilling}
\bibfield{author}{\bibinfo{person}{Nithya Sambasivan} {and}
  \bibinfo{person}{Rajesh Veeraraghavan}.} \bibinfo{year}{2022}\natexlab{}.
\newblock \showarticletitle{The Deskilling of Domain Expertise in AI
  Development}. In \bibinfo{booktitle}{\emph{Proceedings of the Conference on
  Human Factors in Computing Systems}}. \bibinfo{pages}{1--14}.
\newblock


\bibitem[Sarker(2021)]%
        {sarker2021machine}
\bibfield{author}{\bibinfo{person}{Iqbal~H Sarker}.}
  \bibinfo{year}{2021}\natexlab{}.
\newblock \showarticletitle{Machine learning: Algorithms, real-world
  applications and research directions}.
\newblock \bibinfo{journal}{\emph{SN computer science}} \bibinfo{volume}{2},
  \bibinfo{number}{3} (\bibinfo{year}{2021}), \bibinfo{pages}{160}.
\newblock


\bibitem[Schelter et~al\mbox{.}(2018)]%
        {schelter2018automating}
\bibfield{author}{\bibinfo{person}{Sebastian Schelter}, \bibinfo{person}{Dustin
  Lange}, \bibinfo{person}{Philipp Schmidt}, \bibinfo{person}{Meltem Celikel},
  \bibinfo{person}{Felix Biessmann}, {and} \bibinfo{person}{Andreas
  Grafberger}.} \bibinfo{year}{2018}\natexlab{}.
\newblock \showarticletitle{Automating Large-Scale Data Quality Verification}.
\newblock \bibinfo{journal}{\emph{Proceedings of the VLDB Endowment}}
  \bibinfo{volume}{11}, \bibinfo{number}{12} (\bibinfo{year}{2018}),
  \bibinfo{pages}{1781--1794}.
\newblock


\bibitem[Sculley et~al\mbox{.}(2015)]%
        {sculley2015hidden}
\bibfield{author}{\bibinfo{person}{David Sculley}, \bibinfo{person}{Gary Holt},
  \bibinfo{person}{Daniel Golovin}, \bibinfo{person}{Eugene Davydov},
  \bibinfo{person}{Todd Phillips}, \bibinfo{person}{Dietmar Ebner},
  \bibinfo{person}{Vinay Chaudhary}, \bibinfo{person}{Michael Young},
  \bibinfo{person}{Jean-Francois Crespo}, {and} \bibinfo{person}{Dan
  Dennison}.} \bibinfo{year}{2015}\natexlab{}.
\newblock \showarticletitle{Hidden technical debt in machine learning systems}.
\newblock \bibinfo{journal}{\emph{Advances in neural information processing
  systems}}  \bibinfo{volume}{28} (\bibinfo{year}{2015}).
\newblock


\bibitem[Serban et~al\mbox{.}(2020)]%
        {serban2020adoption}
\bibfield{author}{\bibinfo{person}{Alex Serban}, \bibinfo{person}{Koen van~der
  Blom}, \bibinfo{person}{Holger Hoos}, {and} \bibinfo{person}{Joost Visser}.}
  \bibinfo{year}{2020}\natexlab{}.
\newblock \showarticletitle{Adoption and effects of software engineering best
  practices in machine learning}. In \bibinfo{booktitle}{\emph{Proceedings of
  the 14th ACM/IEEE International Symposium on Empirical Software Engineering
  and Measurement (ESEM)}}. \bibinfo{pages}{1--12}.
\newblock


\bibitem[Serban and Visser(2022)]%
        {serban2022adapting}
\bibfield{author}{\bibinfo{person}{Alex Serban} {and} \bibinfo{person}{Joost
  Visser}.} \bibinfo{year}{2022}\natexlab{}.
\newblock \showarticletitle{Adapting software architectures to machine learning
  challenges}. In \bibinfo{booktitle}{\emph{2022 IEEE International Conference
  on Software Analysis, Evolution and Reengineering (SANER)}}. IEEE,
  \bibinfo{pages}{152--163}.
\newblock


\bibitem[Shaw and Zhu(2022)]%
        {shaw2022can}
\bibfield{author}{\bibinfo{person}{Mary Shaw} {and} \bibinfo{person}{Liming
  Zhu}.} \bibinfo{year}{2022}\natexlab{}.
\newblock \showarticletitle{Can software engineering harness the benefits of
  advanced AI?}
\newblock \bibinfo{journal}{\emph{IEEE Software}} \bibinfo{volume}{39},
  \bibinfo{number}{6} (\bibinfo{year}{2022}), \bibinfo{pages}{99--104}.
\newblock


\bibitem[Siddik and Bezemer(2023)]%
        {siddik2023code}
\bibfield{author}{\bibinfo{person}{Md~Saeed Siddik} {and}
  \bibinfo{person}{Cor-Paul Bezemer}.} \bibinfo{year}{2023}\natexlab{}.
\newblock \showarticletitle{Do Code Quality and Style Issues Differ Across
  (Non-) Machine Learning Notebooks? Yes!}. In \bibinfo{booktitle}{\emph{2023
  IEEE 23rd International Working Conference on Source Code Analysis and
  Manipulation (SCAM)}}. IEEE, \bibinfo{pages}{72--83}.
\newblock


\bibitem[Smith et~al\mbox{.}(2020)]%
        {smith2020machine}
\bibfield{author}{\bibinfo{person}{Micah~J Smith}, \bibinfo{person}{Carles
  Sala}, \bibinfo{person}{James~Max Kanter}, {and} \bibinfo{person}{Kalyan
  Veeramachaneni}.} \bibinfo{year}{2020}\natexlab{}.
\newblock \showarticletitle{The machine learning bazaar: Harnessing the ml
  ecosystem for effective system development}. In
  \bibinfo{booktitle}{\emph{Proceedings of the 2020 ACM SIGMOD International
  Conference on Management of Data}}. \bibinfo{pages}{785--800}.
\newblock


\bibitem[Studer et~al\mbox{.}(2021)]%
        {studer2021towards}
\bibfield{author}{\bibinfo{person}{Stefan Studer}, \bibinfo{person}{Thanh~Binh
  Bui}, \bibinfo{person}{Christian Drescher}, \bibinfo{person}{Alexander
  Hanuschkin}, \bibinfo{person}{Ludwig Winkler}, \bibinfo{person}{Steven
  Peters}, {and} \bibinfo{person}{Klaus-Robert M{\"u}ller}.}
  \bibinfo{year}{2021}\natexlab{}.
\newblock \showarticletitle{Towards CRISP-ML (Q): a machine learning process
  model with quality assurance methodology}.
\newblock \bibinfo{journal}{\emph{Machine learning and knowledge extraction}}
  \bibinfo{volume}{3}, \bibinfo{number}{2} (\bibinfo{year}{2021}),
  \bibinfo{pages}{392--413}.
\newblock


\bibitem[Vogelsang and Borg(2019)]%
        {vogelsang2019requirements}
\bibfield{author}{\bibinfo{person}{Andreas Vogelsang} {and}
  \bibinfo{person}{Markus Borg}.} \bibinfo{year}{2019}\natexlab{}.
\newblock \showarticletitle{Requirements engineering for machine learning:
  Perspectives from data scientists}. In \bibinfo{booktitle}{\emph{2019 IEEE
  27th International Requirements Engineering Conference Workshops (REW)}}.
  IEEE, \bibinfo{pages}{245--251}.
\newblock


\bibitem[Walden et~al\mbox{.}(2015)]%
        {walden2015systems}
\bibfield{author}{\bibinfo{person}{David~D Walden} {et~al\mbox{.}}}
  \bibinfo{year}{2015}\natexlab{}.
\newblock \showarticletitle{Systems engineering handbook: A guide for system
  life cycle processes and activities}.
\newblock  (\bibinfo{year}{2015}).
\newblock


\bibitem[Wan et~al\mbox{.}(2019)]%
        {wan2019does}
\bibfield{author}{\bibinfo{person}{Zhiyuan Wan}, \bibinfo{person}{Xin Xia},
  \bibinfo{person}{David Lo}, {and} \bibinfo{person}{Gail~C Murphy}.}
  \bibinfo{year}{2019}\natexlab{}.
\newblock \showarticletitle{How does machine learning change software
  development practices?}
\newblock \bibinfo{journal}{\emph{IEEE Transactions on Software Engineering}}
  \bibinfo{volume}{47}, \bibinfo{number}{9} (\bibinfo{year}{2019}),
  \bibinfo{pages}{1857--1871}.
\newblock


\bibitem[Wasson(2006)]%
        {Wasson-SystemsEngineeringCopingComplexity}
\bibfield{author}{\bibinfo{person}{Charles~S. Wasson}.}
  \bibinfo{year}{2006}\natexlab{}.
\newblock \bibinfo{booktitle}{\emph{Systems Engineering: Coping with
  Complexity}}.
\newblock \bibinfo{publisher}{John Wiley \& Sons}.
\newblock


\bibitem[Whang and Lee(2020)]%
        {whang2020data}
\bibfield{author}{\bibinfo{person}{Steven~Euijong Whang} {and}
  \bibinfo{person}{Jae-Gil Lee}.} \bibinfo{year}{2020}\natexlab{}.
\newblock \showarticletitle{Data collection and quality challenges for deep
  learning}.
\newblock \bibinfo{journal}{\emph{Proceedings of the VLDB Endowment}}
  \bibinfo{volume}{13}, \bibinfo{number}{12} (\bibinfo{year}{2020}),
  \bibinfo{pages}{3429--3432}.
\newblock


\bibitem[Wikipedia(2023)]%
        {wiki:Systems_engineering}
\bibfield{author}{\bibinfo{person}{Wikipedia}.}
  \bibinfo{year}{2023}\natexlab{}.
\newblock \bibinfo{title}{{Systems engineering} --- {W}ikipedia{,} The Free
  Encyclopedia}.
\newblock
  \bibinfo{howpublished}{\url{http://en.wikipedia.org/w/index.php?title=Systems\%20engineering&oldid=1163252030}}.
\newblock
\newblock
\shownote{[Online; accessed 08-August-2023]}.


\bibitem[Wilhjelm and Younis(2020)]%
        {wilhjelm2020threat}
\bibfield{author}{\bibinfo{person}{Carl Wilhjelm} {and}
  \bibinfo{person}{Awad~A. Younis}.} \bibinfo{year}{2020}\natexlab{}.
\newblock \showarticletitle{A threat analysis methodology for security
  requirements elicitation in machine learning based systems}. In
  \bibinfo{booktitle}{\emph{2020 IEEE 20th International Conference on Software
  Quality, Reliability and Security Companion (QRS-C)}}. IEEE,
  \bibinfo{pages}{426--433}.
\newblock


\bibitem[Wu et~al\mbox{.}(2023)]%
        {wu2023comparison}
\bibfield{author}{\bibinfo{person}{Jie~JW Wu}, \bibinfo{person}{Thomas~A
  Mazzuchi}, {and} \bibinfo{person}{Shahram Sarkani}.}
  \bibinfo{year}{2023}\natexlab{}.
\newblock \showarticletitle{Comparison of multi-criteria decision-making
  methods for online controlled experiments in a launch decision-making
  framework}.
\newblock \bibinfo{journal}{\emph{Information and Software Technology}}
  \bibinfo{volume}{155} (\bibinfo{year}{2023}), \bibinfo{pages}{107115}.
\newblock


\bibitem[Zendel et~al\mbox{.}(2015)]%
        {zendel2015cvhazop}
\bibfield{author}{\bibinfo{person}{Oliver Zendel}, \bibinfo{person}{Markus
  Murschitz}, \bibinfo{person}{Martin Humenberger}, {and}
  \bibinfo{person}{Wolfgang Herzner}.} \bibinfo{year}{2015}\natexlab{}.
\newblock \showarticletitle{CV-HAZOP: Introducing test data validation for
  computer vision}. In \bibinfo{booktitle}{\emph{Proceedings of the IEEE
  International Conference on Computer Vision}}. \bibinfo{pages}{2066--2074}.
\newblock


\bibitem[Zhang et~al\mbox{.}(2020)]%
        {zhang2020machine}
\bibfield{author}{\bibinfo{person}{Jie~M. Zhang}, \bibinfo{person}{Mark
  Harman}, \bibinfo{person}{Lei Ma}, {and} \bibinfo{person}{Yang Liu}.}
  \bibinfo{year}{2020}\natexlab{}.
\newblock \showarticletitle{Machine learning testing: Survey, landscapes and
  horizons}.
\newblock \bibinfo{journal}{\emph{IEEE Transactions on Software Engineering}}
  (\bibinfo{year}{2020}).
\newblock


\end{thebibliography}

\end{document}